\documentclass[prb,aps,10pt,twocolumn,showpacs,floats,amsmath,amssymb,amsfonts,mathtools,groupedaddress]{revtex4-1}

\usepackage{hyperref}
\usepackage{graphicx}
\usepackage{dcolumn}
\usepackage{bm}
\usepackage[latin1]{inputenc}
\usepackage{subfigure}
\usepackage{color} 
\usepackage{xcolor}
\usepackage{pifont,wasysym}
\usepackage{enumerate}
\usepackage{enumitem}
\usepackage{physics}

\definecolor{orange}{rgb}{1,0.5,0}
\definecolor{darkgreen}{rgb}{0.13, 0.55, 0.13}

\newcommand\commentout[1]{}

\makeatletter
\newcommand{\mathleft}{\@fleqntrue\@mathmargin0pt}
\newcommand{\mathcenter}{\@fleqnfalse}
\makeatother

\newcommand{\iu}{{i\mkern1mu}} 	
\newcommand{\dint}{\mathrm{d}} 	

\newcommand{\imag}{\mathrm{Im}}
\newcommand{\ie}{{\it i.e.~}} 	

\begin{document}

\title{Simulating time-dependent thermoelectric transport in quantum systems}

\author{Adel Kara Slimane}
\author{Phillipp Reck}
\author{Genevi\`eve Fleury}
\email{genevieve.fleury@cea.fr}
\affiliation{Universit\'e Paris-Saclay, CEA, CNRS, SPEC, 91191 Gif-sur-Yvette, France}

\begin{abstract}
We put forward a gauge-invariant theoretical framework for studying time-resolved thermoelectric transport in an arbitrary multiterminal electronic quantum system described by a noninteracting tight-binding model. The system is driven out of equilibrium by an external time-dependent electromagnetic field (switched on at time $t_0$) and possibly by static temperature or electrochemical potential biases applied (from the remote past) between the electronic reservoirs. 
Numerical simulations are conducted by extending to energy transport 
the wave-function approach developed by Gaury \textit{et al.} and implemented in the t-Kwant library. We provide a module that allows us to compute the time-resolved heat currents and powers in addition to the (already implemented) charge currents, and thus to simulate dynamical thermoelectric transport through realistic devices, when electron-electron and electron-phonon interactions can be neglected. We apply our method to the noninteracting Resonant Level Model and verify that we recover the results reported in the literature for the time-resolved heat currents in the expected limits. Finally, we showcase the versatility of the library by simulating dynamical thermal transport in a Quantum Point Contact subjected to voltage pulses. 
\end{abstract}


\maketitle

\section{Introduction}
Since its early stages in the 1960s,\cite{tien1963} research in time-dependent quantum nanoelectronics has predominantly focused on charge transport. Milestone achievements in the AC regime include \textit{e.g.} the realization of electron pumps,\cite{Pothier1992,giazotto2011} the measurement of the relaxation times of RC\cite{gabelli2006} and LC\cite{gabelli2007} quantum circuits, the observation of radiative signatures of dynamical Coulomb Blockade,\cite{hofheinz2011} and the dynamical measurement of the fractional charge of anyons.\cite{Kapfer2019} In the last decade, the experimental realization of single-electron sources\cite{feve2007,Dubois2013,kataoka2016} has opened up a new research avenue in time-dependent nanoelectronics,\cite{Bauerle2018} with potential applications to quantum computing. On the theory side, the usual frameworks to handle time-dependent transport in quantum electronic systems are the Non Equilibrium Green's Function (NEGF) approach\cite{jauho1994} and the time-dependent scattering formalism,\cite{blanter2000} combined with the Floquet theory\cite{moskalets2011,kohler2005} for time-periodic perturbations. In practice, the NEGF equations are extremely difficult to integrate, even numerically,\cite{prociuk2008,croy2009,wang2011} so that alternative computational strategies have been developed.\cite{cini1980,stefanucci2005,stefanucci2008,bokes2008,xie2012,krueckl2013} In particular, a novel wave-function based approach\cite{gaury2014a,weston2016a,weston2016b} implemented in the t-Kwant library\cite{tKwant} has recently made possible the simulation of time-resolved quantum transport in realistic mesoscopic devices.\cite{gaury2014b,gaury2015,gaury2016,abbout2018,rossignol2019} \\
\indent Dynamical charge transport has thus been the subject of an
intense experimental and theoretical activity in the last decades. In comparison, the study of energy and heat transport in time-dependent quantum electron systems is an emerging research topic. Experimental investigations in the field are challenging and currently at their infancy.\cite{Pekola2015}  A breakthrough has been achieved recently by Karimi \textit{et al.}\cite{Karimi2019} with the measurement of heat and temperature fluctuations in superconducting quantum circuits. Nevertheless, the literature in the field is largely dominated by theoretical works, in particular those studying mesoscopic systems with periodic driving.\cite{ludovico2016bis} They can be roughly ranged into three categories: A first one investigating the fundamentals of quantum thermodynamics,\cite{moskalets2014,ludovico2014,ludovico2016,whitney2018,dou2018} a second one assessing the applicative potential of high-frequency nanoelectronics for AC-driven thermoelectrics\cite{crepieux2011,lim2013,chirla2014,zhou2015, dare2016, gallego2017,ma1999}, heat pumping,\cite{Moskalets2002,rey2007,Pekola2007,arrachea2007,haupt2013} or Josephson-effect-based refrigeration,\cite{solinas2016,virtanen2017} and a third one analyzing energy current and noise as new probes of mesoscopic electron systems.\cite{battista2014, dashti2019} From a technical point of view, a  wide range of approaches have been pursued to deal with dynamical energy and heat transport in mesoscopic electron systems subjected to time-dependent bias and gate voltages, \textit{e.g.} the Floquet theory in the AC regime,\cite{battista2014,moskalets2014,ludovico2014,ludovico2016,ludovico2016bis,gallego2017} the master equation approach\cite{Torfason2013,haupt2013,kosloff2013} often assuming slow driving and weak system-reservoir coupling, the well-established (but cumbersome) NEGF technique,\cite{arrachea2007,crepieux2011, Liu2012, esposito2015, zhou2015, dare2016,yu2016} and more recently the wave-function\cite{michelini2019} and the auxiliary-mode \cite{Lehmann2018} approaches. The effect of Coulomb interaction has been included within different frameworks, near the adiabatic regime\cite{lim2013,romero2017,haupt2013,dou2018} and beyond.\cite{chirla2014,chen2015,whitney2018} Interestingly, alternative methods have also been developed to describe transient particle and heat currents in response to the application of a temperature gradient.\cite{biele2015,eich2016,covito2018,lozej2018}  However, to date, the different methods listed above have only been applied to paradigmatic systems ranging mostly from the single site Resonant Level Model (RLM) to the one-dimensional chain.\\
\indent In addition to the technical difficulty of solving the time-dependent quantum problem, the study of energy transport and conversion in open electronic quantum systems driven by time-dependent potentials is hindered by fundamental issues about the nonequilibrium thermodynamical description of such systems. In particular, the question of the proper definition of a time-dependent heat current (in consistency with thermodynamic requirements) is still under debate. A major difficulty arises from the ill-defined splitting\cite{bruch2016,ochoa2016} between the central system and the electronic reservoirs in the strong coupling regime. Using the time-dependent RLM as a prototypical model, it was argued in Refs.\cite{ludovico2014,ludovico2016} that half of the contribution coming from the energy stored in the system-reservoir coupling region should be included in the definition of the heat current flowing into a given reservoir. This result, derived in the case where the time dependence is restricted to the central region, was later questioned in Refs.\cite{esposito2015,esposito2015bis} when the system-reservoir coupling is also made time-dependent. It was finally generalized to the case of time-dependent coupling by the inclusion of an extra term in the heat current definition.\cite{haughian2018} Recently, Bruch \textit{et al.}\cite{bruch2018} developed an alternative approach based on the Landauer-B\"uttiker scattering theory that circumvents the problem of the system-reservoir splitting.\\ 
\indent In the present paper, we put forward a general framework for the simulation of time-resolved thermoelectric transport in realistic mesoscopic devices subjected to external time-dependent electromagnetic fields. 
The system under consideration is made of an arbitrary (noninteracting) electron scattering region coupled through ideal (noninteracting) leads to electronic reservoirs at local equilibrium. Spin is not included. Till a given time $t_0$, the tight-binding Hamiltonian describing the system is considered as time-independent but the system is possibly driven in a nonequilibrium steady state by the application of (static) electrochemical potential or temperature gradients between the reservoirs. After $t_0$, an external time-dependent electromagnetic field is applied. It may account for the presence of voltage pulses in the leads, time-varying electrostatic gates in the vicinity of the electron gas, or time-dependent magnetic fields in the scattering region. 
Our first task consists of building a thermoelectric framework that involves quantities (particle, energy, and power densities, particle and energy currents) that are all invariant under an arbitrary gauge transformation of the external electromagnetic field. For that purpose, we follow Refs.\cite{yang1976,kobe1982,kobe1987} and define an energy operator that differs in general from the Hamiltonian operator, since the expectation value of the Hamiltonian is generally not gauge invariant. Secondly, to compute the different quantities numerically in an efficient way, we leverage the wave-function approach developed in Refs.\cite{gaury2014a,weston2016a,weston2016b} for the simulation of time-dependent quantum transport. Till recently,\cite{michelini2019} this approach had only been considered within the context of charge transport.\cite{gaury2014b,gaury2015,gaury2016,abbout2018,rossignol2019} Its generalization to energy transport was addressed in Ref.\cite{michelini2019} for the study of a molecular network model. Here, we formulate a general gauge-invariant framework for simulating time-dependent thermoelectric transport in an arbitrary (noninteracting) electron system. We report on its practical implementation as a thermoelectric extension of the t-Kwant simulation library, discuss how it converges to the usual Landauer-B\"uttiker approach in the static limit, and check the validity of our approach by using the Resonant Level Model (RLM) as a test bed. A short investigation of time-dependent heat transport in a Quantum Point Contact (QPC) is also provided for illustrating the potential of our t-Kwant based numerical tool. However, little emphasis is put in this article on physical interpretations for specific examples. This is left for future works. Our approach which inherits the benefits of t-Kwant brings within reach the simulation of time-resolved heat and thermoelectric transport in large realistic systems. It can handle arbitrary time-dependent perturbations, beyond the single-frequency AC limit and the adiabatic regime. Moreover, it does not rely on the wide-band limit hypothesis which is commonly assumed in works using the NEGF technique.\\ 
\indent The outline of the paper is as follows. We define our general noninteracting and time-dependent tight-binding model in Sec.\,\ref{sec:model}. Then in Sec.\,\ref{sec:currentsdf}, we draw up a gauge-invariant thermoelectric framework in terms of the lesser Green's function
of the system. The numerical method used to calculate the time-dependent (particle, energy, and heat) currents as well as the time-dependent powers is introduced in Sec.\,\ref{sec_numericalmethod}. It is based on the wave-function approach developed in Refs.\cite{gaury2014a,weston2016a,weston2016b} which draws upon a reformulation of the NEGF equations in terms of the time-dependent scattering states.\commentout{and which turns out to be well suited for a numerical implementation.} We review briefly the approach and explain how to use it for energy transport. In Sec.\ref{sec:RLMbenchmark}, we apply our method to the Resonant Level (toy) Model and benchmark our results against the ones obtained with other techniques in previous works. We show that we reproduce the NEGF results in the wide-band limit and the results of Ref.\cite{covito2018} beyond this limit. Finally in Sec.\ref{sec:QPC}, we demonstrate the feasibility of large scale simulations by computing time-dependent heat currents in a QPC subjected to a temperature gradient and to a voltage pulse.
We conclude in Sec.\ref{sec:ccl} and discuss briefly possible continuations of this work.

\section{Tight-binding model}
\label{sec:model}
We model noninteracting spinless electrons in an open system made of a scattering region $\mathcal{S}$ connected to an arbitrary number $M$ of semi-infinite leads $\mathcal{L}_\alpha$ ($\alpha=1$ to $M$). The system is discretized on a lattice (with lattice spacing $a=1$) and from the remote past till a given time $t_0$, it is described by the general quadratic Hamiltonian 
\begin{equation}
    \hat{H}(t\leq t_0)=\hat{H}^0=\sum_{i,j}H_{ij}^0\hat{c}_i^\dagger \hat{c}_j
\end{equation}
$\hat{c}_i^\dagger$ [resp. $\hat{c}_i$] being the creation [resp. annihilation] operator of an electron on site $i$ at position $\mathbf{r}_i$. The leads have a translation-invariant structure made up of an infinite repetition of interconnected identical unit cells and hopping terms between two different leads are set to zero. Importantly, the couplings between the (first cell of the) leads and the scattering region is included in the static Hamiltonian $\hat{H}^0$. For this reason, our approach is similar to the so-called partition-free approach.\cite{cini1980, stefanucci2005, stefanucci2008, eich2016} $\hat{H}^0$ also accounts for the presence of any static electromagnetic fields due e.g. to surrounding metallic gates or to the application of voltage biases between leads. Finally, each lead $\mathcal{L}_\alpha$ is attached to a reservoir in thermodynamic equilibrium characterized by an electrochemical potential $\mu_\alpha$ and a temperature $T_\alpha$.\\
\indent For times $t$ larger than $t_0$, an external time-dependent electromagnetic field is applied
\begin{subequations}
\begin{align}
    \mathbf{E}(\mathbf{r},t)&=-\mathbf{\nabla}V(\mathbf{r},t)-\frac{\partial \mathbf{A}}{\partial t}(\mathbf{r},t) \label{eq_Efield}\\
    \mathbf{B}(\mathbf{r},t)&=\mathbf{\nabla}\times\mathbf{A}(\mathbf{r},t)
\end{align}
\end{subequations}
$V(\mathbf{r},t)$ and $\mathbf{A}(\mathbf{r},t)$ being the electromagnetic scalar and vector potentials at position $\mathbf{r}$ at time $t$. 
The system Hamiltonian becomes
\begin{equation}
    \label{eq_H1}
    \hat{H}(t> t_0)=\sum_{i,j}H_{ij}(t)\hat{c}_i^\dagger \hat{c}_j
\end{equation}
with
\begin{equation}
    H_{ij}(t)=H_{ij}^0+H'_{ij}(t)\,.
\end{equation}
The Hamiltonian $\hat{H}'(t)=\sum_{i,j}H'_{ij}(t)\hat{c}_i^\dagger \hat{c}_j$ accounts for the presence of the external time-dependent electromagnetic field when $t> t_0$.
In the scattering region, the field may be fully arbitrary and we have
\begin{equation}
    \label{eq_Hprime}
   H'_{ij}(t)=eV_{i}(t)\delta_{ij}+H_{ij}^0(e^{i\phi_{ij}(t)}-1)(1-\delta_{ij})
\end{equation}
when $i$ and $j$ both lie in $\mathcal{S}$ (or one of them lies in the first cell of one lead). In Eq.\eqref{eq_Hprime}, $e$ denotes the electron~charge, $V_{i}(t)=V(\mathbf{r}_i,t)$ and $\phi_{ij}(t)=(e/\hbar)\int_{\mathbf{r}_j}^{\mathbf{r}_i}\!\mathbf{A}(\mathbf{r},t)\cdot\mathrm{d}\mathbf{r}$ is a Peierls phase\footnote{It is assumed that $\mathbf{A}$ does not vary much on the scale of $\mathbf{r}_j-\mathbf{r}_i$ when employing the Peierls substitution\cite{luttinger1951}} accounting for $\mathbf{A}(\mathbf{r},t)$. In the leads $\mathcal{L}_\alpha$, no time-dependent magnetic fields but only \textit{homogeneous} time-dependent electric potentials $V_i(t)=V_\alpha(t)$ are supposed to be applied so that
\begin{equation}
   \label{eq_H4}
   H'_{ij}(t)=eV_\alpha(t)\delta_{ij}~~~(i\in \mathcal{L}_\alpha, j\in \mathcal{L}_\alpha)\,.
\end{equation}
If needed, the abrupt drop of $V_i(t)$ at the interface between the leads and the scattering region $\mathcal{S}$ can be easily absorbed by including the first cells of the leads into $\mathcal{S}$ and by defining an effective\footnote{In principle, a proper treatment of the electrostatics would require a self-consistent approach. This point has been addressed in Ref.\cite{kloss2018} in the context of L{\"u}ttinger liquids.} screened potential $V_i(t)$ in the vicinity of the interface. Besides, in an actual device, a time-dependent voltage source may induce a variation of the chemical potential in the electronic reservoir, in addition to a variation of the electric potential. This case is not handled in the present paper as it would require modeling relaxation inside the reservoirs. Yet a qualitative discussion of the role of the electrostatics in realistic devices (reported in Sec.8.4 of Ref.\cite{gaury2014a}) shows that the above model has broad applicability in the field of time-dependent quantum nanoelectronics.

\section{Gauge-invariant time-resolved thermoelectric framework}
\label{sec:currentsdf}
Here we construct the theoretical framework that will be used to describe time-dependent thermoelectric transport in the generic tight-binding model defined above. Our aim is to formulate a theory that is invariant under any gauge transformation of the external electromagnetic field. For that purpose, we follow Refs.\cite{yang1976,kobe1982,kobe1987} and define an energy operator that differs in general from the Hamiltonian operator. We start this section with a succinct reminder of gauge invariance in quantum mechanics, then we draw our theoretical picture of time-dependent thermoelectrics from the local to the global scale.

\subsection{Gauge transformations}

We consider an arbitrary local gauge transformation of the electromagnetic field 
\begin{subequations}
\begin{align}
    V(\mathbf{r}_i,t) \rightarrow \widetilde{V}(\mathbf{r}_i,t) & =V(\mathbf{r}_i,t)-\frac{\partial \Lambda(\mathbf{r}_i,t)}{\partial t} \\
    \mathbf{A}(\mathbf{r}_i,t) \rightarrow \widetilde{\mathbf{A}}(\mathbf{r}_i,t) & =\mathbf{A}(\mathbf{r}_i,t)+\mathbf{\nabla}\Lambda(\mathbf{r}_i,t)
\end{align}
\label{eq_gaugetransfo}
\end{subequations}
where $\Lambda(\mathbf{r}_i,t)$ is an arbitrary, differentiable, real function of space and time. Under Eq.\eqref{eq_gaugetransfo}, the Hamiltonian $\hat{H}(t)$ transforms as 
\begin{align}
    \hat{H}(t) \rightarrow \hat{\widetilde{H}}(t)=\sum_{i,j}\widetilde{H}_{ij}(t)\hat{c}_i^\dagger \hat{c}_j
\end{align}
where
\begin{subequations}
\label{eq_Htilde}
\begin{align}
    \widetilde{H}_{ii}(t)&=H_{ii}(t)-e\frac{\partial \Lambda_i(t)}{\partial t} \label{eq_Hiitilde} \\
    \widetilde{H}_{ij}(t)&=H_{ij}(t)e^{i\frac{e}{\hbar}[\Lambda_i(t)-\Lambda_j(t)]}~~(\mathrm{if}~i\neq j) \label{eq_Hijtilde}
\end{align}
\end{subequations}
and $\Lambda_i(t)=\Lambda(\mathbf{r}_i,t)$. In particular, for $t\leq t_0$ when $V(\mathbf{r}_i,t)=0$ and $\mathbf{A}(\mathbf{r}_i,t)=\mathbf{0}$, the gauge-transformed static Hamiltonian $\widetilde{H}(t\leq t_0)=\widetilde{H}^0$ may become artificially time-dependent. In the rest of the paper, we fix the gauge when $t \leq t_0$ and no time-dependent electromagnetic field is applied. We choose the natural gauge in which $\Lambda_i(t\leq t_0)=0$.\\ 
\indent The electromagnetic gauge transformation \eqref{eq_gaugetransfo} can also be understood\,\cite{kobe1979}  as a change of basis of the one-body orbitals on sites $i$ associated with the operators $\hat{c}_i$.
Under Eq.\,\eqref{eq_gaugetransfo} a unitary transformation $\hat{U}(t)=\exp\left(i\frac{e}{\hbar}\sum_i\Lambda_i(t)\hat{c}_i^\dagger \hat{c}_i\right)$ is made on the annihilation operator
\begin{equation}
    \label{eq_citilde}
    \hat{c}_i \rightarrow \hat{\widetilde{c}}_i=\hat{U}\hat{c}_i\hat{U}^\dagger=e^{-i\frac{e}{\hbar}\Lambda_i(t)}c_i\,
\end{equation}
so that the transformed Hamiltonian $\hat{\widetilde{H}}$ can be written as 
\begin{equation}
\label{eq_gaugeonH}
    \hat{\widetilde{H}}=\hat{U} \hat{H} \hat{U}^\dagger -i\hbar\,\hat{U} \frac{\partial \hat{U}^\dagger}{\partial t}
\end{equation}
after having noticed that
\begin{equation}
    \label{eq_UdrUdagger}
    i\hbar\,\hat{U} \frac{\partial \hat{U}^\dagger}{\partial t}=e\sum_i \frac{\partial \Lambda_i}{\partial t}\hat{c}_i^\dagger \hat{c}_i=e\sum_{i} (V_i-\widetilde{V}_i)\hat{c}_i^\dagger \hat{c}_i\,.
\end{equation}
The Schr\"odinger equation
\begin{equation}
    i\hbar\frac{\partial |\Psi(t)\rangle}{\partial t}=\hat{H}(t)|\Psi(t)\rangle
\end{equation}
written here for an arbitrary solution $|\Psi(t)\rangle$, turns out to be invariant in form under the local gauge transformation \eqref{eq_gaugetransfo}
\begin{equation}
    i\hbar\frac{\partial |\widetilde{\Psi}(t)\rangle}{\partial t}=\hat{\widetilde{H}}(t)|\widetilde{\Psi}(t)\rangle
\end{equation}
since the wave function $|\Psi(t)\rangle$ transforms as 
\begin{equation}
\label{eq_gaugeonPsi}
    |\Psi(t)\rangle \rightarrow |\widetilde{\Psi}(t)\rangle=U|\Psi(t)\rangle\,.
\end{equation}
While any Hermitian operator $\hat{O}=\sum_{i,j}O_{ij}\hat{c}_i^\dagger \hat{c}_j$ that transforms as
\begin{equation}
     \hat{O} \rightarrow \hat{\widetilde{O}}=\hat{U} \hat{O} \hat{U}^\dagger
\end{equation}
under Eq.\eqref{eq_gaugetransfo} has a gauge invariant expectation value
\begin{equation}
    \langle \Psi| \hat{O} | \Psi \rangle = \langle \widetilde{\Psi}| \hat{\widetilde{O}} | \widetilde{\Psi} \rangle
\end{equation}
the Hamiltonian does not (see Eq. (\ref{eq_gaugeonH})): its expectation value is in general not gauge invariant
\begin{equation}
    \langle \Psi| \hat{H}| \Psi \rangle \neq \langle \widetilde{\Psi}| \hat{\widetilde{H}} | \widetilde{\Psi} \rangle
\end{equation}
and thus it cannot be considered as the energy operator.\footnote{The same conclusion holds also in classical mechanics when an external time-dependent electromagnetic field is present (see Ref.\cite{kobe1987}).}

\subsection{Local quantities}
In this section, we define the energy density and the local energy current carried by electrons inside the scattering region. We also derive the related continuity equation. It involves the local power injected in the system by the time-dependent external electromagnetic field. The different quantities are constructed following Refs.\cite{yang1976,kobe1982,kobe1987} so as to be gauge invariant. They are formally written in terms of the lesser Green's function of the system. The continuity equation for the electron density is also recalled for completeness. 

\subsubsection{Particle density and local particle current}
We introduce the lesser Green's function
\begin{equation}
    \label{eq_lessergreendf}
    G_{ij}^<(t,t') =\frac{i}{\hbar} \langle \hat{c}_j^\dagger(t') \hat{c}_i(t) \rangle
\end{equation}
where $\hat{c}_i(t)$ is the annihilation operator in the Heisenberg representation. The change over time of the electron density
\begin{equation}
    \rho^N_i (t)= \langle \hat{c}_i^\dagger(t) \hat{c}_i(t) \rangle = -i\hbar G_{ii}^<(t,t)
    \label{eq_rhoN}
\end{equation}
can be calculated by using the equation of motion
\begin{equation}
    \label{eq_ofmotion}
     \frac{\dint \hat{O}_\mathcal{H}}{\dint t} =  \left.\frac{\partial \hat{O}}{\partial t}\right|_\mathcal{H} + \frac{\iu}{\hbar}[\hat{H}_\mathcal{H},\hat{O}_\mathcal{H}]
\end{equation}
here written for an arbitrary operator $\hat{O}$. $\hat{O}_\mathcal{H}$ and $\hat{H}_\mathcal{H}=\sum_{i,j}H_{ij}(t)\hat{c}_i^\dagger(t) \hat{c}_j(t)$ correspond to $\hat{O}$ and $\hat{H}$ in the Heisenberg representation. The subscript $\mathcal{H}$ will be dropped in the rest of the paper. We find the continuity equation for the particle number 
\begin{equation}
    \label{eq_continuityparticle}
    \frac{\mathrm{d}\rho^N_i}{\mathrm{d}t}+\sum_{j\neq i}I^N_{ji}(t)=0
\end{equation}
where 
\begin{equation}
    I^N_{ij}(t) =2\,\mathrm{Re} \left[ H_{ji}(t)G_{ij}^<(t,t)\right ]
    \label{eq:I^N_loc}
\end{equation}
is the local particle current between sites $i$ and $j$, that satisfies $I^N_{ji}=-I^N_{ij}$. Using Eqs.\eqref{eq_Hijtilde}, \eqref{eq_citilde} and \eqref{eq_gaugeonPsi}, it is easy to check that $\rho^N_i$ and $I^N_{ij}$ are gauge independent quantities.

\subsubsection{Energy density, local energy current, and power density}
\label{sec:localenergy}
In the static case, the energy operator coincides with the Hamiltonian operator. Actually, as pointed out in Ref.\cite{kobe1987}, this is only true if the scalar and vector potentials used to describe the static electromagnetic field are taken time independent but it is customary and natural to choose such a gauge for static fields. 

In the time-dependent case, we follow Refs.\cite{yang1976,kobe1982,kobe1987} and define the energy operator $\hat{\varepsilon}$ as
\begin{equation}
\label{eq_kobeoperator}
\hat{\varepsilon}(t)=\hat{H}(t)-e\hat{V}(t)
\end{equation}
where $\hat{V}(t)=\sum_i V_i(t)\hat{c}_i^\dagger \hat{c}_i$. The sum is made here over all sites $i$ in the whole system, including the leads $\mathcal{L}_\alpha$ where $V_i=V_\alpha$. 
By construction, $\langle \hat{\varepsilon} \rangle$ is gauge invariant. Indeed
\begin{equation}
\langle \Psi| \hat{H}-e\hat{V} | \Psi \rangle = \langle \widetilde{\Psi}| \hat{\widetilde{H}}-e\hat{\widetilde{V}} | \widetilde{\Psi} \rangle   
\end{equation}
in virtue of Eqs.\,\eqref{eq_gaugeonH}, \eqref{eq_UdrUdagger}, and \eqref{eq_gaugeonPsi}.

With this definition, the energy operator reads
$\hat{\varepsilon}(t)=\sum_{i,j}\varepsilon_{ij}(t)\hat{c}_i^\dagger \hat{c}_j$ with
\begin{subequations}
\label{eq_kobeenergycoeff}
\begin{align}
    \varepsilon_{ii}(t) &= H_{ii}^0 \\
    \varepsilon_{ij}(t) &= H_{ij}(t)~~~\mathrm{if}~i\neq j
\end{align}
\end{subequations}
and the gauge transformed energy operator reads $\hat{\widetilde{\varepsilon}}(t)=\sum_{i,j}\widetilde{\varepsilon}_{ij}(t)\hat{c}_i^\dagger \hat{c}_j$
with
\begin{subequations}
\label{eq_kobeenergycoefftilde}
\begin{align}
    \widetilde{\varepsilon}_{ii}(t) &= H_{ii}^0 \\
    \widetilde{\varepsilon}_{ij}(t) &= H_{ij}(t)e^{i\frac{e}{\hbar}[\Lambda_i(t)-\Lambda_j(t)]}~~~\mathrm{if}~i\neq j\,.
\end{align}
\end{subequations}

We now write the energy operator $\hat{\varepsilon}$ as a sum of local energy density operators $\hat{\varepsilon}_i$ 
\begin{equation}
    \label{eq:sumofEi}
    \hat{\varepsilon}=\sum_i  \hat{\varepsilon}_i
\end{equation}
with
\begin{equation}
    \label{eq_energydensity}
    \hat{\varepsilon}_i(t)=\varepsilon_{ii}(t)\hat{c}_i^\dagger \hat{c}_i+\frac{1}{2}\sum_{j\neq i}\left(\varepsilon_{ij}(t)\hat{c}_i^\dagger \hat{c}_j+\varepsilon_{ji}(t)\hat{c}_j^\dagger \hat{c}_i\right)
\end{equation}
containing both the on-site potential energy $\varepsilon_{ii}$ on site $i$ and half of the kinetic energy $\varepsilon_{ij}$ stored in the hoppings. Note that the definition of the energy density operator is not unique: There exist other $\hat{\varepsilon}_i$ that yield Eq.\eqref{eq:sumofEi} and there are \textit{a priori} no physical but only technical grounds to favor one choice over the others, as noticed in Ref.\cite{mathews1974}. Our definition \eqref{eq_energydensity} corresponds to the discretized version on a lattice of the energy density of the second kind put forward in Ref.\cite{mathews1974}, which leads to a symmetrized local energy current ($I^E_{ji}$ below). The same definition has been used in \textit{e.g.} Refs.\cite{ludovico2014,michelini2019,Wu_2008}. To be fully precise, Refs.\cite{ludovico2014,michelini2019,mathews1974,Wu_2008} dealt with the  
inherent ambiguity in the definition of the Hamiltonian density operator $\hat{H}_i$ (so as $\hat{H}=\sum_i \hat{H}_i$) but the same discussion holds for $\hat{\varepsilon}_i$.

Let us now introduce $\rho^E_i=\langle \hat{\varepsilon}_i \rangle$ the (gauge invariant) energy density on site $i$ and let us write down the continuity equation for the energy with the help of Eqs.\eqref{eq_lessergreendf}, \eqref{eq_ofmotion}, and \eqref{eq_energydensity}. We find 
\begin{equation}
    \label{eq_continuityenergy}
    \frac{\mathrm{d}\rho^E_i}{\mathrm{d}t}+\sum_{j\neq i}I^E_{ji}(t)=S_i^E(t)
\end{equation} 
where the local energy current $I^E_{ij}$ between sites $i$ and $j$ is given by
\begin{align}
    \label{IEij_via_G}
    I^E_{ij}(t)=\sum_k \mathrm{Re}&\left[\varepsilon_{kj}(t)\varepsilon_{ji}(t)G_{ik}^<(t,t)\right. \nonumber\\ 
    &-\left.\varepsilon_{ki}(t)\varepsilon_{ij}(t)G_{jk}^<(t,t)\right]
\end{align} 
and the power source term $S_i^E$ on site $i$ by 
\begin{align}
\label{eq_sourceterm}
    S_i^E(t)=\sum_{j\neq i}&\left\lbrace\hbar\,\mathrm{Im}\left[\frac{\partial \varepsilon_{ij}}{\partial t}G_{ji}^<(t,t)\right]\right. \nonumber\\ 
    &+ \mathrm{Re}\left[eV_i(t)\varepsilon_{ij}(t)G_{ji}^<(t,t)\right.  \nonumber\\
    &~~~~~~~~~~+\left.\left.eV_j(t)\varepsilon_{ji}(t)G_{ij}^<(t,t)\right]\right\rbrace\,.
\end{align} 
Using Eqs.\eqref{eq:I^N_loc} and \eqref{eq_kobeenergycoeff}, the source term can be written as
\begin{equation}
    \label{eq_sourceterm2}
    S_i^E(t)=\sum_{j\neq i}\left[ \frac{1}{2}\left(V_i(t)-V_j(t)\right)+\frac{\hbar}{2e}\frac{\partial \phi_{ij}}{\partial t}\right]eI^N_{ji}(t)
\end{equation}
which is the discretized expression on a grid of the electric power density $e\,\mathbf{j}(\mathbf{r}_i,t)\cdot\mathbf{E}(\mathbf{r}_i,t)$, $\mathbf{j}(\mathbf{r}_i,t)$ being the particle current density and  $\mathbf{E}(\mathbf{r}_i,t)$ the external electric field given by Eq.\eqref{eq_Efield}. Eqs.\eqref{eq_continuityenergy} to \eqref{eq_sourceterm2} are derived in Appendix \ref{app:proof_eqcontinuity}. It is worth noting that the splitting into local energy current $I^E_{ji}(t)$ and source term $S_i^E(t)$ from the time derivative of the energy density in the continuity equation \eqref{eq_continuityenergy} is not unique. However the choice made above has several advantages:
\begin{itemize}[label=$\ast$]
    \item $I^E_{ji}$ and $S_i$ are gauge invariant quantities.
    \item The energy change over the full system is given by only the external power source and not the energy currents, \textit{i.e.}
\begin{equation}
     \sum\limits_i \frac{\mathrm{d}\rho^E_i}{\mathrm{d}t} = \sum\limits_i S^E_i(t)\,.
\end{equation}
    \item The source term $S_i$ coincides with the electric power density supplied by the external time-dependent electromagnetic field.
    \item $I^E_{ji}=-I^E_{ij}$ to fit the interpretation of a net current flowing from $i$ to $j$. Moreover $I^E_{ji}=0$ if the sites $i$ and $j\neq i$ are disconnected (\textit{i.e.} if $H_{ji}=0$). The identification of a local energy current $I^E_{ij}$ out of the divergence $ \sum_{j\neq i} I^E_{ij}$ is however not unequivocal. Another choice $\mathcal{I}^E_{ij}\neq I^E_{ij}$ was made in Ref.\cite{michelini2019}. It satisfies $\sum_j\mathcal{I}^E_{ij}= \sum_j I^E_{ij}$ and $\mathcal{I}^E_{ji}=-\mathcal{I}^E_{ij}$ but contrary to $I^E_{ij}$, $\mathcal{I}^E_{ij}$ can be non-zero between two disconnected sites.
    \item In the static case $t\leq t_0$, the local energy current integrated over a lead $\mathcal{L}_\alpha$ (\textit{i.e} the energy current flowing in a lead  $\mathcal{L}_\alpha$) coincides with the usual static definition of the energy current.\cite{Butcher1990,benenti2017} This will be discussed in Section \ref{sec:staticlim}.
\end{itemize}

\subsection{Global currents}
\label{sec_globalcurrent}

Hereafter, we integrate over space the local quantities introduced above and define in particular the particle, energy, and heat currents flowing in the leads. In the limit where the external electromagnetic field becomes time-independent, we compare those currents to the static ones given by the usual Landauer-B\"uttiker formulas. As the latter formulas are derived by taking the Hamiltonian as the energy operator, the two energy currents are different in general. However the two heat currents coincide in the static limit, as well as the two particle currents. 

\subsubsection{Particle currents}
We define
\begin{subequations}
\begin{align}
    N_\mathcal{S}(t)&=\sum_{i\in\mathcal{S}}\rho_i^N(t)\\
    N_{\alpha}(t)&=\sum_{i\in\mathcal{L}_\alpha}\rho_i^N(t)   \label{eq:df_Nalpha}
\end{align}
\end{subequations}
the particle number at time $t$ in the scattering region $\mathcal{S}$ and in each lead $\mathcal{L}_\alpha$. The continuity equation \eqref{eq_continuityparticle} integrated over space yields
\begin{subequations}
\begin{align}
    I^N_\mathcal{S}(t)&=-\frac{\mathrm{d}N_\mathcal{S}}{\mathrm{d}t}\\
    I^N_{\alpha}(t)&=-\frac{\mathrm{d}N_{\alpha}}{\mathrm{d}t} \label{eq_timedepparticlecurrent}
\end{align}
\end{subequations}
where
\begin{equation}
    I^N_{\alpha}(t)=\sum_{i\in\mathcal{L}_\alpha}\sum_{j\in\mathcal{S}}I^N_{ji}(t)
\end{equation}
is the (incoming) particle current in the lead $\mathcal{L}_\alpha$ (\textit{i.e.} flowing toward the scattering region) and 
\begin{equation}
    \label{eq_IN_lead}
    I^N_\mathcal{S}(t)=\sum_{i\in\mathcal{S}}\sum_\alpha\sum_{j\in\mathcal{L}_\alpha}I^N_{ji}(t)
\end{equation}
the displacement current. The total particle number is conserved and we have
\begin{equation}
    I^N_\mathcal{S}(t)+\sum_{\alpha}I^N_{\alpha}(t)=0\,.
\end{equation}

\subsubsection{Energy currents}
We proceed similarly for the energy-related quantities and define
\begin{subequations}
\begin{align}
    E_\mathcal{S}(t)&=\sum_{i\in\mathcal{S}}\rho_i^E(t)\\
    E_{\alpha}(t)&=\sum_{i\in\mathcal{L}_\alpha}\rho_i^E(t)   \label{eq:df_Ealpha}
\end{align}
\end{subequations}
the energy at time $t$ in the scattering region $\mathcal{S}$ and in each lead $\mathcal{L}_\alpha$. By integrating spatially the continuity equation \eqref{eq_continuityenergy}, we get
\begin{subequations}
\begin{align}
    \frac{\mathrm{d}E_\mathcal{S}}{\mathrm{d}t}+I^E_\mathcal{S}(t)&=S_\mathcal{S}^E(t)\\
    \frac{\mathrm{d}E_{\alpha}}{\mathrm{d}t}+I^E_{\alpha}(t)&=S_\alpha^E(t) \label{eq_globalenergycontinuity}
\end{align}
\end{subequations}
where
\begin{subequations}
\begin{align}
    I^E_\mathcal{S}(t)&=\sum_{i\in\mathcal{S}}\sum_\alpha\sum_{j\in\mathcal{L}_\alpha}I^E_{ji}(t)\\
    I^E_{\alpha}(t)&=\sum_{i\in\mathcal{L}_\alpha}\sum_{j\in\mathcal{S}}I^E_{ji}(t)     \label{eq_timedepenergycurrent}
\end{align}
\end{subequations}
are respectively the energy current in the scattering region and the (incoming) energy current in the lead $\mathcal{L}_\alpha$, while
\begin{subequations}
\begin{align}
    S^E_\mathcal{S}(t)&=\sum_{i\in\mathcal{S}} S_i^E(t)\\
    S^E_\alpha(t)&=\sum_{i\in\mathcal{L}_\alpha} S_i^E(t)
\end{align}
\end{subequations}
are the external power source terms. Using $I^E_{ji}=-I^E_{ij}$, we obtain the following conservation equation
\begin{equation}
    \frac{\mathrm{d}E_\mathcal{S}}{\mathrm{d}t} + \sum_\alpha \frac{\mathrm{d}E_{\alpha}}{\mathrm{d}t}=S_\mathcal{S}^E(t)+\sum_\alpha S_\alpha^E(t)
\end{equation}
\ie the variation of the energy stored in the whole system equals the power supplied by the external electromagnetic field. Note that if for instance $V_i(t)=V_\alpha(t)$ are applied in the leads $\mathcal{L}_\alpha$ but not in the scattering region ($V_i(t)=0$ if $i\in\mathcal{S}$), and if no time-dependent magnetic field is applied, then the source terms reduce to
\begin{subequations}
\label{eq_source_form}
\begin{align}
    S_\alpha^E(t)&=\frac{e}{2}V_\alpha(t)I_\alpha^N(t)\\
    S_\mathcal{S}^E(t)&=\frac{e}{2}\sum_{\alpha} V_\alpha(t)I_\alpha^N(t)
\end{align}
\end{subequations}
and the total source term is $S_\mathcal{S}^E+\sum_\alpha S_\alpha^E=e\sum_\alpha V_\alpha I_\alpha^N$.

\subsubsection{Heat currents}
We define the (incoming) time-dependent heat current in the lead $\mathcal{L}_\alpha$ as
\begin{equation}
    \label{eq_timedepheatcurrent}
    I^H_\alpha(t)=I^E_\alpha(t)-S_\alpha^E(t)-\mu_\alpha I_\alpha^N(t)
\end{equation}
where $\mu_\alpha$ is the electrochemical potential of the reservoir attached to $\mathcal{L}_\alpha$ defined for $t\leq t_0$ in Sec.\,\ref{sec:model}. This definition leads to the following formulation of the first law of thermodynamics
\begin{equation}
    \label{eq_1stlaw}
    \frac{\mathrm{d}E_\mathcal{S}}{\mathrm{d}t}=S_\mathcal{S}^E(t)+\sum_\alpha S_\alpha^E(t)+\sum_\alpha\mu_\alpha I^N_\alpha(t)+\sum_\alpha I^H_\alpha(t)
\end{equation}
where the first three terms on the right hand side of Eq.\eqref{eq_1stlaw} correspond to the rate of work supplied to the scattering region $\mathcal{S}$ by its environment.
$I^H_\alpha(t)$ defined above is gauge invariant since $I^E_\alpha(t)$, $S_\alpha^E(t)$ and $I_\alpha^N(t)$ are. 
In a gauge where the leads are time-independent, the following equality holds (see Appendix \ref{app:heatcurrentdf})
\begin{equation}
    \label{eq_timedepheatcurrent_vs_lit}
    I^H_\alpha(t)=-\frac{\mathrm{d}}{\mathrm{d}t}\langle \hat{\widetilde{H}}_{\alpha}+\frac{1}{2}\hat{\widetilde{H}}_{\mathcal{S}\alpha}\rangle-\mu_\alpha I_\alpha^N(t)
\end{equation}
where $\hat{\widetilde{H}}_{\alpha}$ and $\hat{\widetilde{H}}_{\mathcal{S}\alpha}$ denote respectively the gauge transformed Hamiltonian of the lead $\mathcal{L}_\alpha$ and the gauge transformed tunneling Hamiltonian between $\mathcal{L}_\alpha$ and the scattering region $\mathcal{S}$. Note that $\hat{\widetilde{H}}_{\mathcal{S}\alpha}$ may depend on time but not $\hat{\widetilde{H}}_{\alpha}$ by construction. The term on the right-hand side of Eq.\eqref{eq_timedepheatcurrent_vs_lit} was used to define the lead heat current in Refs.\cite{ludovico2014,ludovico2016} in the case where the time-dependent perturbations are confined to the scattering region. It was argued in Ref.\cite{haughian2018} that an extra term has to be added in the heat current definition when the tunneling Hamiltonian is also time-dependent. However in our case, this extra term is zero\footnote{The extra term has been derived in Ref.\cite{haughian2018} for the RLM under the wide-band limit approximation. It vanishes if the modulus of the hopping terms connecting the dot and the reservoirs is time independent. In the main text, we have implicitly conjectured that this conclusion remains true for a generic tight-binding model beyond the wide-band limit. Strictly speaking, we can only assert that the lead heat current defined by Eq.\eqref{eq_timedepheatcurrent} coincides with the one defined in Ref.\cite{haughian2018} under the hypothesis made in the latter work.} because the time dependency in the hopping terms of our model Hamiltonian only appears through a time-dependent phase, $\widetilde{H}_{ij}(t)=H_{ij}^0e^{i\widetilde{\phi}_{ij}(t)}$ (whose origin is a time-dependent Peierls substitution or gauge transformation). Therefore, Eq.\eqref{eq_timedepheatcurrent} is a gauge invariant formulation of the lead heat current which coincides with the definition used in Refs.\cite{ludovico2014,ludovico2016,haughian2018} when a gauge in which the leads are time-independent is chosen. This point will be discussed in more details in Appendix \ref{app:heatcurrentdf}.
Besides, we will see in the next subsection that $I^H_\alpha(t)$ also coincides with the heat current given by the Landauer-B\"uttiker formula in the particular limit where $\hat{H}(t)$ becomes time-independent. Finally, it is noteworthy that if we include the first cell of each lead into the scattering region, and define thereby new leads $\mathcal{L}_{\bar{\alpha}}$, then $S_{\bar{\alpha}}^E(t)=0$ and we have $I^H_{\bar{\alpha}}(t)=I^E_{\bar{\alpha}}(t)-\mu_\alpha I_{\bar{\alpha}}^N(t)$. 

\subsubsection{Static limit}
\label{sec:staticlim}
When $t\leq t_0$ and no external time-dependent electromagnetic field is applied, the system Hamiltonian is time independent \textit{i.e.} $\hat{H}(t\leq t_0)=\hat{H}^0$. Within the static Landauer-B\"uttiker formalism, the particle, energy, and heat currents in the lead $\mathcal{L}_\alpha$ read\cite{Butcher1990,benenti2017}
\begin{subequations}
\label{eq_Istatic_LB}
\begin{align}
    I_\alpha^{N,st}&=\sum_{\beta \neq \alpha}\int \frac{\mathrm{d}E}{h}\,[f_{\mu_\alpha,T_\alpha}(E)-f_{\mu_\beta,T_\beta}(E)]\,T_{\alpha\beta}(E) \label{eq_IN_static}\\
    I_\alpha^{E,st}&=\sum_{\beta \neq \alpha}\int\frac{\mathrm{d}E}{h}\,[f_{\mu_\alpha,T_\alpha}(E)-f_{\mu_\beta,T_\beta}(E)]\,E\,T_{\alpha\beta}(E) \label{eq_IE_static}\\
    I_\alpha^{H,st}&=I_\alpha^{E,st}-\mu_\alpha I_\alpha^{N,st} \label{eq_IH_static}
\end{align}
\end{subequations}
where $f_{\mu,T}(E)=[1+\exp(\frac{E-\mu}{k_B T})]^{-1}$ is the Fermi function ($k_B$ being the Boltzmann constant), the sum over $\beta$ is a sum over leads $\mathcal{L}_\beta$ and $T_{\alpha\beta}(E)=\sum_{m_\alpha}\sum_{m_\beta}|S^{\alpha\beta}_{m_\alpha m_\beta}(E)|^2$ is the probability for an electron at energy $E$ to be transmitted from the lead $\mathcal{L}_\beta$ into the lead $\mathcal{L}_\alpha$ ($S^{\alpha\beta}_{m_\alpha m_\beta}(E)$ being the scattering amplitude from the mode $m_\beta$ at energy $E$ in $\mathcal{L}_\beta$ to the mode $m_\alpha$ at energy $E$ in $\mathcal{L}_\alpha$). 
It is straightforward to show that the particle $I_\alpha^N$, energy $I_\alpha^E$, and heat $I_\alpha^H$ currents, defined above in Eqs.\eqref{eq_timedepparticlecurrent}, \eqref{eq_timedepenergycurrent}, and \eqref{eq_timedepheatcurrent} respectively, equal the standard static current formulas 
\begin{subequations}
\label{eq_statcurrent1}
\begin{align}
     I_\alpha^N(t\leq t_0) &= I_\alpha^{N,st}\\
     I_\alpha^E(t\leq t_0) &= I_\alpha^{E,st}\\
     I_\alpha^H(t\leq t_0) &= I_\alpha^{H,st}\,.
\end{align}
\end{subequations}
This is true even in the time-dependent gauge (see the comment below Eq.\eqref{eq_Htilde}).

Let us now consider the case for $t>t_0$ where an external time-dependent electromagnetic field is applied and let us assume that the field converges to a static limit at long times \textit{i.e} $\hat{H}(t\to\infty)=\hat{H}^{\bar{st}}$. In this new static configuration, the static particle $I_\alpha^{N,\bar{st}}$, energy $I_\alpha^{E,\bar{st}}$, and heat currents $I_\alpha^{H,\bar{st}}$ are given by the Landauer-B\"uttiker formulas \eqref{eq_Istatic_LB} with $\mu_\alpha \to \mu_\alpha+eV_\alpha$ and $T_{\alpha\beta}\to \bar{T}_{\alpha\beta}$. Here, $V_\alpha \equiv V_\alpha(t\to\infty)$ in the leads $\mathcal{L}_\alpha$ while $\bar{T}_{\alpha\beta}$ denote the transmissions of the system defined by $\hat{H}^{\bar{st}}$. 
In Appendix \ref{app:static}, we show
\begin{subequations}
\label{eq_statcurrent2}
\begin{align}
     I_\alpha^N(t\to \infty) &= I_\alpha^{N,\bar{st}}  \label{eq_INstaticlimit}\\
     I_\alpha^E(t\to \infty) &= I_\alpha^{E,\bar{st}}-eV_\alpha I_\alpha^{N,\bar{st}}+S_\alpha^E(t\to \infty) \label{eq_IEstaticlimit}\\
     I_\alpha^H(t\to \infty) &= I_\alpha^{E,\bar{st}}-(\mu_\alpha+eV_\alpha) I_\alpha^{N,\bar{st}}=I_\alpha^{H,\bar{st}}\,. \label{eq_IHstaticlimit}
\end{align}
\end{subequations}
Thus in the static limit $t\to\infty$, the energy currents $I_\alpha^E(t\to \infty)$ in the leads $\mathcal{L}_\alpha$ differ from the usual static energy currents $I_\alpha^{E,\bar{st}}$. This is due to the fact that $I_\alpha^{E,\bar{st}}$ is calculated by defining the energy operator as $\hat{\varepsilon}=\hat{H}^{\bar{st}}$ while $I_\alpha^E(t\to \infty)$ is calculated using $\hat{\varepsilon}(t\to \infty)=\hat{H}^{\bar{st}}-e\hat{V}(t\to \infty)$ (see Eq.\eqref{eq_kobeoperator}). The discrepancy $I_\alpha^E(t\to \infty) \neq I_\alpha^{E,\bar{st}}$ is the price to pay for a gauge-invariant energy current $I_\alpha^E(t)$ that also satisfies $I_\alpha^E(t\leq t_0) = I_\alpha^{E,st}$. It stems from the definition of $\hat{\varepsilon}$ in Eq.\eqref{eq_kobeoperator} as the sum of the kinetic energy and of the static potential that is present on the system from the remote past. In the peculiar case where the external electromagnetic field converges to a static limit (and then varies again), it might be relevant to redefine the energy operator with respect to this new static configuration and forget the past. More importantly, the heat currents $I_\alpha^H(t)$ which are written as a difference of energy currents are not affected by this choice of the reference static potential. We find that $I_\alpha^H(t)$ converges to the usual static heat current in the static limit (see Eq.\eqref{eq_IHstaticlimit}).

\section{Numerical method}
\label{sec_numericalmethod}
We now discuss how to compute in an efficient way the time-dependent quantities introduced in Sec.\,\ref{sec:currentsdf}. We use for that purpose the wave-function based approach developed in Refs.\cite{gaury2014a,weston2016a,weston2016b} which has been shown to be formally equivalent to the NEGF formalism but much more efficient from a computational point of view. This approach is at the root of the numerical package t-Kwant,\cite{tKwant} which extends to the time domain the quantum transport package Kwant.\cite{groth2014,santos2019,kwant} To date, t-Kwant has been used for calculating time resolved particle density and particle current in various systems.\cite{gaury2014b,gaury2015,abbout2018,kloss2018,rossignol2018,rossignol2019} The case of particle current noise has also been dealt with in Ref.\cite{gaury2016}. Hereafter, we present succinctly the t-Kwant algorithm and explain how to use it for the calculation of the time resolved energy density, power density, energy current and heat current. Note that the same wave-function based approach has recently been used in Ref.\cite{michelini2019} for calculating energy currents in molecular networks. We report here on its implementation in the t-Kwant package.

\subsection{Choice of the electromagnetic gauge}
\label{sec_tkwantgauge}
\commentout{All quantities introduced in Sec.\,\ref{sec:currentsdf} \textit{i.e.} $\rho_i^N$, $\rho_i^E$, $I_{ij}^N$, $I_{ij}^E$, and $S_i^E$ have been shown to be gauge invariant. We are therefore free to choose a convenient electromagnetic gauge for numerical calculation. This choice is guided by two requirements of the t-Kwant algorithm: \textit{(i)} the need for a time-independent system Hamiltonian for $t\leq t_0$, and \textit{(ii)} the need for time-independent leads at all times. To fulfill those two conditions, the gauge function $\Lambda_i(t)$ is chosen as
\begin{subequations}
\label{eq:df_tkwantgauge}
\begin{align}
\Lambda_i(t\leq t_0)&=0~~\mathrm{everywhere}\\
\Lambda_i(t> t_0)&=\phi_\alpha(t)~~~\mathrm{if}~i\in\mathcal{L}_\alpha\\
\Lambda_i(t> t_0)&~~\mathrm{arbitrary}~~~\mathrm{if}~i\in\mathcal{S}
\end{align}
\end{subequations}
where $\phi_\alpha(t)=\int_{t_0}^t \mathrm{d}u\,V_\alpha(u)$. Indeed, under the gauge transformation \eqref{eq_gaugetransfo}, the Hamiltonian defined in Eqs.\eqref{eq_H1}-\eqref{eq_H4} transforms as $\hat{H}(t) \rightarrow \hat{\widetilde{H}}(t)$ according to Eq.\eqref{eq_Htilde}, and one finds eventually time-independent leads in the new gauge
\begin{equation}
    \tilde{H}_{ij}(t)=H_{ij}^0~~~\mathrm{if}~i\,\in\,\mathcal{L}_\alpha, j\,\in\,\mathcal{L}_\alpha
\end{equation}
while the system-lead coupling terms acquire an additional time-dependent phase
\begin{equation}
    \tilde{H}_{ij}(t)=H_{ij}(t) e^{i\frac{e}{\hbar}(\phi_\alpha(t)-\Lambda_j(t))}~~~\mathrm{if}~i\,\in\,\mathcal{L}_\alpha, j\,\in\,\mathcal{S}\,.
\end{equation}
}
All quantities introduced in Sec.\,\ref{sec:currentsdf} \textit{i.e.} $\rho_i^N$, $\rho_i^E$, $I_{ij}^N$, $I_{ij}^E$, and $S_i^E$ have been shown to be gauge invariant. We are therefore free to choose a convenient electromagnetic gauge for numerical calculation. For $t\leq t_0$, the t-Kwant algorithm requires working in the natural gauge in which the system Hamiltonian is time-independent (\textit{i.e.} $\Lambda_i(t\leq t_0)=0$ everywhere). Another prerequisite for t-Kwant is the absence of time dependency in the leads. This can be always achieved by fixing the gauge function $\Lambda_i(t)$  in the leads to
\begin{equation}
\label{eq:df_tkwantgauge}
    \Lambda_i(t)=\phi_\alpha(t)~~~\mathrm{if}~i\in\mathcal{L}_\alpha 
\end{equation}
where $\phi_\alpha(t)=\int_{t_0}^t \mathrm{d}u\,V_\alpha(u)$. Indeed, under the gauge transformation \eqref{eq_gaugetransfo}, the Hamiltonian defined in Eqs.\eqref{eq_H1}-\eqref{eq_H4} transforms as $\hat{H}(t) \rightarrow \hat{\widetilde{H}}(t)$ according to Eq.\eqref{eq_Htilde}, and one finds eventually time-independent leads in the new gauge
\begin{equation}
    \tilde{H}_{ij}(t)=H_{ij}^0~~~\mathrm{if}~i\,\in\,\mathcal{L}_\alpha, j\,\in\,\mathcal{L}_\alpha
\end{equation}
while the system-lead coupling terms acquire an additional time-dependent phase
\begin{equation}
    \tilde{H}_{ij}(t)=H_{ij}(t) e^{i\frac{e}{\hbar}(\phi_\alpha(t)-\Lambda_j(t))}~~~\mathrm{if}~i\,\in\,\mathcal{L}_\alpha, j\,\in\,\mathcal{S}\,.
\end{equation}
There is however no restriction on the gauge function $\Lambda_j(t)$ in the scattering region $\mathcal{S}$ for $t> t_0$. It can be chosen arbitrarily.

\subsection{Main steps of the t-Kwant algorithm}
In this section, we do not present any original result but outline the main steps of the t-Kwant algorithm following Refs.\cite{gaury2014a,weston2016a,weston2016b}.
We use boldface letters to denote matrices, \textit{e.g.} $\mathbf{\widetilde{H}}$ is the matrix whose elements are $\widetilde{H}_{ij}$. Without loss of generality, we also fix $t_0=0$ in order to lighten the equations below. 

The central objects of the t-Kwant numerical technique are the time-dependent scattering states
$\widetilde{\Psi}^{m_\alpha E}(t)$, solutions of the Schr\"odinger equation
\begin{equation}
    \iu \hbar \frac{\partial}{\partial t}\widetilde{\Psi}^{m_\alpha E}(t) = \mathbf{\widetilde{H}}(t) \widetilde{\Psi}^{m_\alpha E}(t)
    \label{eq:tdepSE}
\end{equation}
with the initial condition
\begin{equation}
    \widetilde{\Psi}^{m_\alpha E}(t=0) = \Psi^{m_\alpha E,0}\,.
    \label{eq:tSE-initcond}
\end{equation}
The tildes on top of $\widetilde{\Psi}^{m_\alpha E}$ and $\mathbf{\widetilde{H}}$ are written to remind us that the t-Kwant gauge \eqref{eq:df_tkwantgauge} is used. The stationary scattering states $\Psi^{m_\alpha E,0}$ labeled by their energy $E$ and their incoming mode $m_\alpha$ (in lead $\mathcal{L}_\alpha$) characterize the static problem for $t\leq 0$
\begin{equation}
    \mathbf{H}^0\Psi^{m_\alpha E,0}=E\Psi^{m_\alpha E,0}\,.
\end{equation}
They can be calculated with the Kwant library. Note that while $E$ corresponds to the energy of the stationary wave function $\Psi^{m_\alpha E,0}$, it cannot be interpreted as the energy of the time-evolved scattering state $\widetilde{\Psi}^{m_\alpha E}(t)$ since energy is not conserved in a time-dependent setup. In practice, Eq.\eqref{eq:tdepSE} is numerically intractable as it is defined on the whole (infinite) lattice. To circumvent this problem, a change o variable $\bar{\Psi}^{m_\alpha E}(t)=e^{iEt/\hbar}\widetilde{\Psi}^{m_\alpha E}(t)-\Psi^{m_\alpha E,0}$ is made. The new wave functions $\bar{\Psi}^{m_\alpha E}(t)$ satisfy the Schr\"odinger-like differential equation
\begin{equation}
    \iu \hbar \frac{\partial}{\partial t}\bar{\Psi}^{m_\alpha E}(t) = [\mathbf{\widetilde{H}}(t)-E] \bar{\Psi}^{m_\alpha E}(t)+S^{m_\alpha E}(t)
    \label{eq:tdepSE2}
\end{equation}
with the initial condition
\begin{equation}
    \bar{\Psi}^{m_\alpha E}(t=0) = 0
    \label{eq:tSE-initcond2}
\end{equation}
and an additional source term
\begin{equation}
    S^{m_\alpha E}(t)=[\mathbf{\widetilde{H}}(t)-\mathbf{H}^0]\Psi^{m_\alpha E,0}
    \label{eq:tSE-source}
\end{equation}
which is non zero only in a finite central region since the leads are time independent in the t-Kwant gauge (see Sec.\ref{sec_tkwantgauge}). For this reason and as \textit{(i)} the initial wave function vanishes everywhere and \textit{(ii)} $\bar{\Psi}^{m_\alpha E}(t)$ is composed of outgoing modes only, it is sufficient to solve Eqs.\eqref{eq:tdepSE2}-\eqref{eq:tSE-source} in a finite system around $\mathcal{S}$ \textit{i.e.} it is possible to truncate the leads (after having calculated $\Psi^{m_\alpha E,0}$). It is necessary however to get rid of spurious reflections of outgoing waves on the truncated lead boundaries. This can be done in different ways.\cite{gaury2014a}  The most efficient one consists of adding an imaginary on-site potential $i\Sigma_x$ over the first lead unit cells (labeled $x=1,2,...$ from the scattering region) and to make it vary smoothly with $x$ in order to absorb the outgoing waves and suppress reflections. Eventually, one solves
\begin{equation}
    \iu \hbar \frac{\partial}{\partial t}\bar{\Psi}^{m_\alpha E}(t) = [\mathbf{\widetilde{H}}(t)-E-i\mathbf{\Sigma}] \bar{\Psi}^{m_\alpha E}(t)+S^{m_\alpha E}(t)
    \label{eq:tdepSE3}
\end{equation}
together with Eqs.\eqref{eq:tSE-initcond2}-\eqref{eq:tSE-source} in a finite region made of $\mathcal{S}$ and a finite portion of the leads. This is done with the Dormand-Prince (Runge-Kutta) method. The sink term which reads $\mathbf{\Sigma}=\Sigma_x \mathbf{1}_{cell}$ in the first lead cells (and $\mathbf{\Sigma}=\mathbf{0}$ elsewhere) can be designed in such a way that the reflection amplitude of outgoing waves is arbitrarily close to zero. Its expression is given in Ref.\cite{weston2016a}, together with more details about the source-sink algorithm outlined above.\\
\indent Once the time-dependent wave functions $\widetilde{\Psi}^{m_\alpha E}(t)$ are computed, the particle density $\rho_i^N(t)$ and the particle current $I^N_{ij}(t)$ given by Eqs.\eqref{eq_rhoN} and \eqref{eq:I^N_loc} respectively can be deduced using\cite{gaury2014a} 
\begin{equation}
 G^<_{ij}(t,t^\prime)  = \iu \sum\limits_\alpha \sum_{m_\alpha} \int\frac{\dint E}{h} f_\alpha(E)\left[\widetilde{\Psi}^{m_\alpha E}_j(t^\prime)\right]^*  \widetilde{\Psi}^{m_\alpha E}_i(t) 
 \label{eq:link_G_WF}
\end{equation}
where $f_\alpha(E)=f_{\mu_\alpha,T_\alpha}(E)$ is a shorthand notation for the Fermi function and $\widetilde{\Psi}^{m_\alpha E}_i(t)$ denotes the wave-function value at site $i$. Eq.\eqref{eq:link_G_WF} is the cornerstone of the present paper as it relates the NEGF approach (used in Sec.\ref{sec:currentsdf}) to the t-Kwant wave-function approach (used hereafter for numerical implementation).
$\rho_i^N(t)=-i\hbar G_{ii}^<(t,t)$ is directly given by Eq.\eqref{eq:link_G_WF} while $I^N_{ij}(t)$ reads in terms of the scattering states
\begin{align}
 I^N_{ij}(t) = -2  &\sum\limits_\alpha \sum_{m_\alpha} \int\frac{\dint E}{h} f_\alpha(E) \nonumber \\
 &\times \imag\left[\left(\widetilde{\Psi}^{m_\alpha E}_j(t)\right)^* \widetilde{H}_{ji}(t) \widetilde{\Psi}^{m_\alpha E}_i(t)\right] .
 \label{eq:IN(t)_via_WF}
\end{align}
Hence both quantities can be computed with t-Kwant by integrating over the scattering states which were initially occupied at $t=0$. In practice, the integration is preferably done in momentum instead of energy (to avoid divergent behavior of the integrand in the vicinity of band openings). We emphasize that Eq.\eqref{eq:IN(t)_via_WF} can also be written without the tildes since $I^N_{ij}(t)$ is gauge invariant. The same goes for $\rho_i^N(t)=-i\hbar G_{ii}^<(t,t)$. However, technically, the calculations are done in the t-Kwant gauge. This is the reason why we kept the tildes in the formulas.\\ 
\indent We end this introductory section of t-Kwant with a short discussion of the performance of this numerical approach. The first version of the algorithm (extensively described in Ref.\cite{gaury2014a}) has been much improved by the inclusion of the (customized) sink term $\mathbf{\Sigma}$ in Eq.\eqref{eq:tdepSE3} (see Ref.\cite{weston2016a}). The computational complexity associated with the calculation of the time-dependent scattering states finally reduces to $\mathcal{O}(Nt_{max})$ where $N$ is the number of sites in the system and $t_{max}$ the maximal time to which wave functions are evolved. This complexity has to be multiplied by $N_E$, the number of points in energy needed for calculating the integral in Eq.\eqref{eq:IN(t)_via_WF}. Typically $20<N_E<100$. In the end it turns out that in terms of computation times, the wave-function-based t-Kwant algorithm outperforms the NEGF-based approaches by several orders of magnitude (see Table I in Ref.\cite{gaury2014a}) though the two formalisms are formally equivalent. This makes possible the simulation of large realistic devices (made of tens of thousands of sites) at simulation times which are long enough to capture the full time-dependent response. Hereafter, we show how to leverage the t-Kwant algorithm for the simulation of time-dependent energy transport.

\subsection{Generalization to energy transport}
The local energy density $\rho^E_i=\langle \hat{\varepsilon}_i \rangle$ and the local energy currents $ I_{ij}^{E}(t)$ written as a function of the lesser Green's function $G^<_{ij}(t,t)$ in Eqs.\eqref{eq_energydensity} and \eqref{IEij_via_G} respectively can be readily expressed in the wave-function formalism with the help of Eq.\eqref{eq:link_G_WF}. One finds\footnote{Eqs.\eqref{eq:rho_E_via_WF} and \eqref{eq:IE(t)_via_WF} for $\rho_i^{E}$ and $I_{ij}^{E}$, as well as their counterparts for $\rho_i^{N}$ and $I_{ij}^{N}$, can be derived in the wave-function formalism, without invoking the lesser Green's function. The proof goes in two steps. We start from the one-body problem in the continuum\cite{mathews1974} and discretize on a lattice the probablity density $\rho^N(\mathbf{r},t)=\psi^*\psi$, the probability current density $\mathbf{j}^N(\mathbf{r},t)=\mathrm{Re}[\psi^*\mathbf{v}\psi]$, the energy density $\rho^E(\mathbf{r},t)=\mathrm{Re}[\psi^*\varepsilon\psi]$, and the (symmetrized) energy current density $\mathbf{j}^E(\mathbf{r},t)=\frac{1}{2}\mathrm{Re}[(\varepsilon\psi)^*(\mathbf{v}\psi)+\psi^*\mathbf{v}\varepsilon \psi]$, where $\psi(\mathbf{r},t)$ is the particle wave function, $\mathbf{v}=(-i\hbar\mathbf{\nabla}-e\mathbf{A})/m$ the velocity operator and $\varepsilon(\mathbf{r},\mathbf{v},t)$ the energy operator of the one-body continuous problem. Those quantities satisfy the continuity equations $\mathrm{d}\rho^N/\mathrm{d}t+\mathbf{\nabla}\cdot\mathbf{j}^{N}=0$ and $\mathrm{d}\rho^E/\mathrm{d}t+\mathbf{\nabla}\cdot\mathbf{j}^{E}=e\,\mathbf{j}^{N}\!\cdot\!\mathbf{E}$. Thereby we obtain the one-body contributions of $\rho_i^{N}$, $I_{ij}^{N}$, $\rho_i^{E}$, and $I_{ij}^{E}$ by identifying $\mathbf{\nabla}\cdot\mathbf{j}^{N[E]}$ with the one-body part of $\sum_{j\neq i}I_{ji}^{N[E]}$. Then, the full $\rho_i^{N}$, $I_{ij}^{N}$, $\rho_i^{E}$, and $I_{ij}^{E}$ corresponding to the many-body (noninteracting) problem are deduced by filling up the one-body states according to the Fermi-Dirac statistics.
} 
\begin{align}
    \label{eq:rho_E_via_WF}
    \rho_i^{E}(t)=& \sum_\alpha\sum_{m_\alpha} \int\frac{\mathrm{d}E}{2\pi} f_\alpha(E) \nonumber \\
    & \times \sum_j\mathrm{Re}\left[ \left(\widetilde{\Psi}_i^{m_\alpha E}(t)\right)^* \widetilde{\varepsilon}_{ij}(t) \,\widetilde{\Psi}_j^{m_\alpha E}(t)\right] 
\end{align}
and
\begin{align}
    \label{eq:IE(t)_via_WF}
    I_{ij}^{E}(t)=& \sum_\alpha\sum_{m_\alpha} \int\frac{\mathrm{d}E}{h} f_\alpha(E) \nonumber \\
    & \times \sum_k\mathrm{Im}\left[ \left(\widetilde{\Psi}_k^{m_\alpha E}(t)\right)^* \widetilde{\varepsilon}_{ki}(t) \widetilde{\varepsilon}_{ij}(t) \,\widetilde{\Psi}_j^{m_\alpha E}(t)\right. \nonumber\\
    &~~~~- \left.\left(\widetilde{\Psi}_k^{m_\alpha E}(t)\right)^* \widetilde{\varepsilon}_{kj}(t) \widetilde{\varepsilon}_{ji}(t) \,\widetilde{\Psi}_i^{m_\alpha E}(t)\right]\,.
\end{align}
Both quantities can be computed with t-Kwant, in the same spirit as $\rho_i^N(t)$ and  $I^N_{ij}(t)$ but with an additional sum over the system sites. The electric power density $S_i^E(t)$ given by Eq.\eqref{eq_sourceterm2} can be computed as well. As before, all tildes can be dropped in Eqs.\eqref{eq:rho_E_via_WF} and \eqref{eq:IE(t)_via_WF} but in practice, the calculation is done in the t-Kwant gauge \textit{i.e} with the tilded quantities. Those local quantities can eventually be summed up over space to deduce for instance the lead energy currents $I^E_\alpha(t)$ and the lead heat currents $I^H_\alpha(t)$. We have implemented an additional Python package \texttt{tkwantoperator} \cite{energytKwant} as an extension to the t-Kwant package \cite{tKwant} to compute these quantities and have shown that the extra CPU time needed for computing these quantities is small in comparison to the time needed for calculating the scattering states (see Appendix \ref{app:performance} for more details). In the following, we perform t-Kwant simulations of (electronic) heat transport in the paradigmatic time-dependent RLM, in order to validate our approach and our numerical implementation. We also report on an exploratory investigation of time-dependent heat transport in a QPC driven by voltage pulses. Without discussing deeply the physics involved, we illustrate the strong potential of the t-Kwant (extended) platform for the study of dynamical thermoelectrics and caloritronics.

\section{Resonant Level Model\\
as a benchmark}
\label{sec:RLMbenchmark}

\commentout{In Refs.\cite{crepieux2011, Liu2012, zhou2015, dare2016}, the NEGF equations were solved in the wide-band limit for the time-dependent Resonant Level (toy) Model mimicking a quantum dot coupled to two electronic reservoirs. An unexplained increase of the Seebeck coefficient\cite{crepieux2011} and of the thermoelectric efficiency\cite{zhou2015} were predicted in the transient regime.}
The (noninteracting) time-dependent RLM has been extensively studied in the literature to simulate dynamical charge transport (see \textit{e.g.} Refs.\cite{jauho1994,Platero2004,ridley2015}) and more recently dynamical energy transport,\cite{crepieux2011, Liu2012, esposito2015, ludovico2016, ludovico2016bis, zhou2015, dare2016, yu2016, Lehmann2018, covito2018, esposito2010} in a single level quantum dot or molecular junction connected to two electronic reservoirs. Hereafter we use this model as a test bed to benchmark our numerical approach described above. We consider two cases: \textit{(i)} when (only) the dot level $\epsilon_0(t)$ is varied in time as $\epsilon_0(t)=\epsilon_0+eV_0\Theta(t)$, $\Theta$ being the Heaviside function, and \textit{(ii)} when the time-dependent step-like perturbation is performed in the leads. We calculate the time-dependent energy and heat currents with our numerical approach and show that we reproduce in the expected limits the results obtained previously in the literature.

\subsection{Model}
\label{sec:RLM_Model}
We consider a one-dimensional (1D) chain made of a central site $0$ with on-site energy $\epsilon_0(t)$ connected through a nearest-neighbor hopping term $\gamma_c$ to two semi-infinite left ($L$, on sites $i\leq -1$) and right ($R$, on sites $i\geq 1$) leads with on-site energies $\epsilon_L(t)$ and $\epsilon_R(t)$, and a nearest-neighbor hopping term $\gamma$. The Hamiltonian reads
\begin{equation}
    \label{eq_RLM_1D_1}
    \hat{H}(t)=\hat{H}_0(t)+\sum_{\alpha=L,R}\hat{H}_\alpha(t)+\sum_{\alpha=L,R}\hat{H}_{0\alpha}
\end{equation}
where
\begin{equation}
    \label{eq_dotHamiltonian}
    \hat{H}_0(t)=\epsilon_0(t)\hat{c}_0^\dagger \hat{c}_0
\end{equation}
is the dot Hamiltonian,
\begin{equation}
    \label{eq_H_RLM_lead}
    \hat{H}_\alpha(t)=\sum_{\pm i\geq 1}\left[\epsilon_\alpha (t)\hat{c}_i^\dagger \hat{c}_{i}+\gamma \,\hat{c}_{i\pm 1}^\dagger \hat{c}_{i}+\gamma \,\hat{c}_{i}^\dagger \hat{c}_{i\pm 1}\right]
\end{equation}
the Hamiltonian of the lead $\alpha=L$ or $R$, and
\begin{equation}
    \label{eq_Hc_RLM}
    \hat{H}_{0\alpha} =\gamma_c \,\hat{c}_0^\dagger \hat{c}_{\pm 1}+h.c.
\end{equation}
the tunneling Hamiltonian between the dot and the lead $\alpha$. In Eqs.\eqref{eq_H_RLM_lead} and \eqref{eq_Hc_RLM}, a $-$ [$+$] sign has to be taken if $\alpha=L$ [$R$]. The parameters $\epsilon_0(t)$ and $\epsilon_\alpha(t)$ are constant in time for $t\leq t_0(=0)$. Note that within the t-Kwant approach, the time-dependence of the lead on-site energies $\epsilon_\alpha(t)=\epsilon_\alpha+eV_\alpha(t)\Theta(t)$ is gauged out ($\epsilon_\alpha(t)\to\epsilon_\alpha$) while the dot-lead hopping term $\gamma_c$ acquires a dynamical phase ($\gamma_c\to\gamma_c e^{-i\frac{e}{\hbar}\int_0^t \mathrm{d}uV_\alpha(u)}$, see Sec.\ref{sec_tkwantgauge}). Finally, each lead $\alpha$ is attached from the remote past to an electronic reservoir at equilibrium with static electrochemical potential $\mu_\alpha$ and temperature $T_\alpha$ defined for $t\leq 0$. They remain at equilibrium for $t>0$. Only the electric part of the electrochemical potential may become time-dependent (depending on the gauge). The chemical potential and the temperature are supposed to remain constant.

\begin{figure}[h!]
    \centering
    \includegraphics[keepaspectratio,width=0.9\columnwidth]{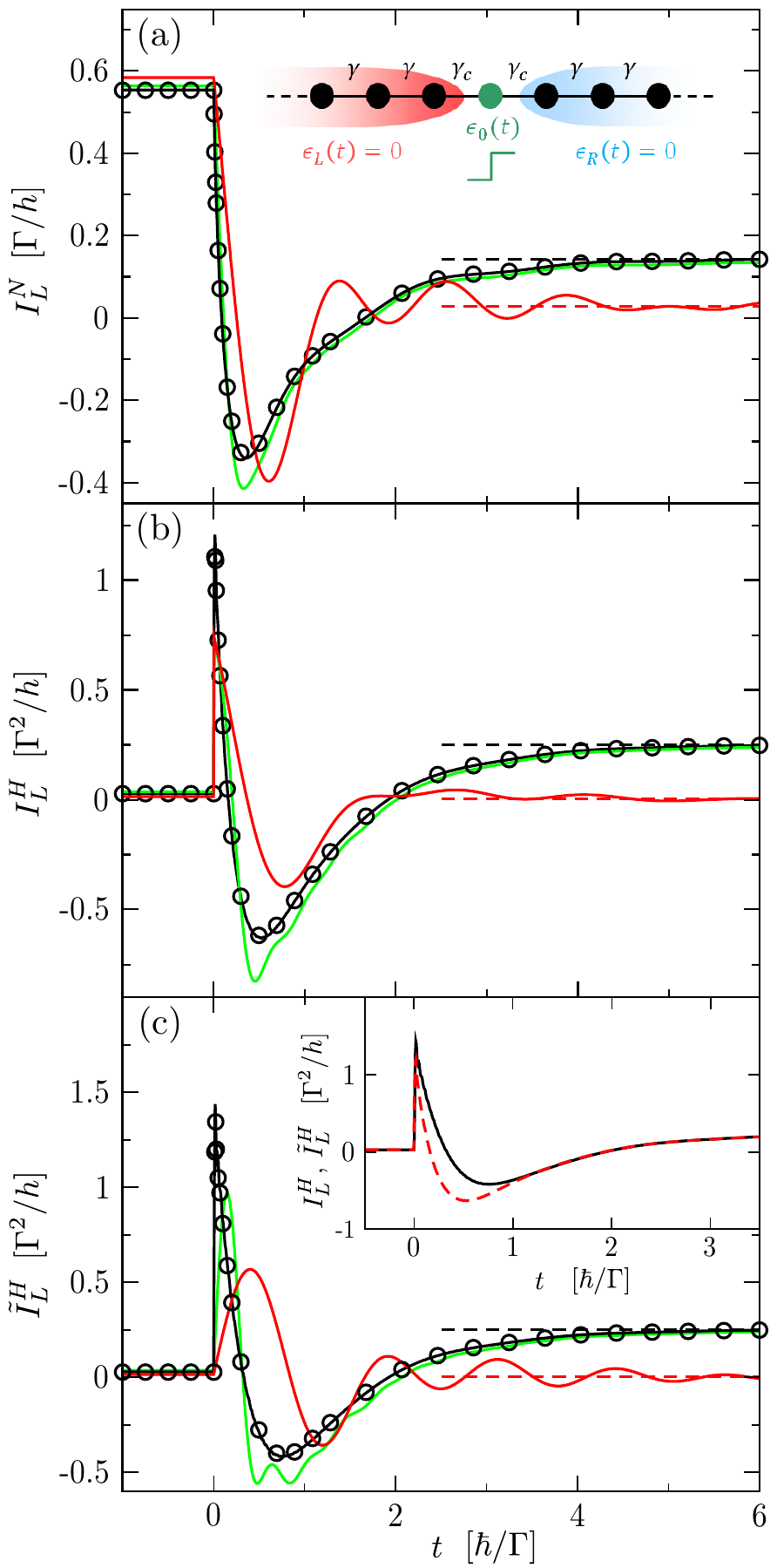}
    \caption{Left particle current $I^N_L$ (a) and left heat currents $I^H_L$ (b) and $\tilde{I}^H_L$ (c) as a function of time $t$, for the 1D~RLM defined by Eqs.\eqref{eq_RLM_1D_1}-\eqref{eq_Hc_RLM}, when the dot energy level is modified as $\epsilon_0(t)=\epsilon_0+eV_0\Theta(t)$ (inset of panel (a)). Units of $x$ and $y$ axes are indicated in brackets. In all panels, data are computed numerically with t-Kwant for different values of $\lambda \gamma/\Gamma$ ($1$ (red lines), $6.25$ (green lines), and $100$ (black lines). The horizontal dashed lines plotted for $\lambda \gamma/\Gamma=1$ (in red) and $100$ (in black) correspond to the static limits at large times $\Gamma t/\hbar\gg 1$ given by the Laudauer-B\"uttiker formulas (see Sec.\ref{sec:staticlim}). When $\lambda \gamma/\Gamma \gg 1$, the t-Kwant results converge to the NEGF results (circles) derived in the wide-band limit (Appendix \ref{app_RLM_NEGF}). Inset of panel (c): comparison of $I^H_L(t)$ (red dashed line) and $\tilde{I}^H_L(t)$ (black line) in the wide-band limit. In all panels, $\epsilon_0=0.5\Gamma$, $eV_0=2.5\Gamma$, $\epsilon_L(t)=\epsilon_R(t)=0$, $T_L=\Gamma/k_B$, $T_R=0$, $\mu_L=0.5\Gamma$, and $\mu_R=-0.5\Gamma$. The NEGF curves are independent of $\Gamma$. The t-Kwant curves are functions of $\lambda \gamma/\Gamma$ and not of the three parameters $\lambda$, $\gamma$, and $\Gamma$ taken separately.}
    \label{fig:RLM_TimeDepInDot}
\end{figure}

\subsection{Results for a time-dependent dot energy level}
\label{sec:RLM_timeindot}

Let us first consider the case where $\epsilon_L(t)=\epsilon_R(t)=0$ while a step-like variation $\epsilon_0(t)=\epsilon_0+eV_0\Theta(t)$ of the dot energy level is applied (see Inset of Fig.\ref{fig:RLM_TimeDepInDot}(a)). This configuration has the advantage of being analytically tractable with the NEGF technique in the so-called wide-band limit approximation. Moreover, since the time-dependent perturbations are restricted to the dot level $\epsilon_0(t)$, the energy operator $\hat{\varepsilon}$ coincides in this case with the Hamiltonian operator in the leads, and we have $E_\alpha=\langle \hat{H}_{\alpha}+\frac{1}{2}\hat{H}_{0\alpha}\rangle$ where $E_\alpha$ is defined by Eq.\eqref{eq:df_Ealpha}. Hereafter, we calculate with t-Kwant the time-dependent heat current $I^H_L(t)=-\frac{\mathrm{d}E_{L}}{\mathrm{d}t}-\mu_L I^N_L(t)$ in \textit{e.g.} the left lead (see Eqs.\eqref{eq_timedepheatcurrent} and \eqref{eq_globalenergycontinuity}) and compare it to the one obtained within the NEGF formalism under the wide-band limit approximation (see Appendix \ref{app_RLM_NEGF}). A similar comparison is done for the particle current $I^N_L(t)$ and for an alternative heat current $\tilde{I}^H_L(t)\equiv -\frac{\mathrm{d}\langle \hat{H}_{L}\rangle}{\mathrm{d}t}-\mu_L I^N_L(t)$ which does not include the contribution of the lead-dot tunneling Hamiltonian $\hat{H}_{0L}$. Such a definition of the heat current was considered in \textit{e.g.} Refs.\cite{crepieux2011,zhou2015}.
Note that in the wave-function formalism, we have for the present model
\begin{align}
    \label{eq:IE(t)_RLM_noIc}
    \frac{\mathrm{d}\langle \hat{H}_{L}\rangle}{\mathrm{d}t}=\,& 2\sum_\alpha\sum_{m_\alpha} \int\frac{\mathrm{d}E}{h} f_\alpha(E) \nonumber \\
    & \times \gamma\gamma_c\,\mathrm{Im}\left[ \left(\Psi_{-2}^{m_\alpha E}(t)\right)^* \,\Psi_{0}^{m_\alpha E}(t)\right]\,. 
\end{align}
This allow us to compute $\tilde{I}^H_L(t)$ with t-Kwant. $I^N_L(t)=I^N_{0,-1}(t)$ and $I^H_L(t)=I^E_{0,-1}(t)-S^E_{-1}(t)-\mu_L I^N_{0,-1}(t)$ are calculated using Eqs.\eqref{eq_sourceterm2}, \eqref{eq:IN(t)_via_WF}, and \eqref{eq:IE(t)_via_WF}.

To make the comparison between the t-Kwant and the NEGF results in the wide-band limit, we follow the scaling approach used in Ref.\cite{covito2018}. We vary simultaneously the hopping terms in the chain by replacing the $\gamma$ and $\gamma_c$ parameters with
\begin{subequations}
\begin{align}
    \bar{\gamma}&=\lambda\gamma  \\
    \bar{\gamma}_c&=\sqrt{\lambda}\gamma_c 
\end{align}
\end{subequations}
where $\lambda$ is a scaling factor. When $\lambda$ is increased, the width $[-2\bar{\gamma},2\bar{\gamma}]$ of the (single) conduction band in the leads widens while the ratio $\Gamma\equiv 4\bar{\gamma}_c^2/\bar{\gamma}$ remains fixed. In the limit $\lambda \gamma/\Gamma\to\infty$ (keeping $\Gamma$ finite), the retarded self-energy $\Sigma^R(E)$ of the (identical) time-independent left and right leads,
\begin{equation}
    \Sigma^R(E)=\frac{\bar{\gamma}_c^2}{\bar{\gamma}}\left[\frac{E}{2\bar{\gamma}}-i\sqrt{1-\left(\frac{E}{2\bar{\gamma}}\right)^2}\right]~\mathrm{if}~\frac{|E|}{2|\bar{\gamma}|}\leq 1\,,
\end{equation}
converges to $-i\frac{\Gamma}{4}$ \textit{i.e.} the real part of $\Sigma^R(E)$ becomes zero and its imaginary part becomes energy independent. This corresponds to the wide-band limit hypothesis.

In Fig.\ref{fig:RLM_TimeDepInDot}, we plot $I^N_L(t)$, $I^H_L(t)$, and $\tilde{I}^H_L(t)$ calculated with t-Kwant for various values of the ratio $\lambda \gamma/\Gamma=\lambda(\gamma/\gamma_c)^2/4$, keeping  the other parameters fixed.\footnote{The set of fixed parameters corresponds to the one used in Ref.\cite{crepieux2011}.} We check that in the wide-band limit $\lambda \gamma/\Gamma \gg 1$, the t-Kwant results (black lines in Fig.\ref{fig:RLM_TimeDepInDot}) converge to the NEGF results\footnote{Note that we did not investigate in details the behavior of the NEGF data at small $t\gtrsim 0$. While the t-Kwant heat currents are observed to be continuous at $t=0$, the NEGF heat currents calculated by integrating numerically Eq.\eqref{eq_app_RLM_integral} (with standard routines) turn out be numerically unstable in the vicinity of $t\gtrsim 0$. This is probably an (irrelevant) artifact of the wide-band limit approximation that leads in some cases to pathological singularities, as pointed out in Ref.\cite{covito2018}.} given in Appendix \ref{app_RLM_NEGF} (circles in Fig.\ref{fig:RLM_TimeDepInDot}). 
Moreover, in the inset of Fig.\ref{fig:RLM_TimeDepInDot}(c), we compare $I^H_L(t)$ and $\tilde{I}^H_L(t)$ and show that both quantities coincide in the long time limit $\Gamma t/\hbar \to \infty$. This is illustrated in the wide-band limit $\lambda \gamma/\Gamma\to\infty$ but holds for any value of $\lambda \gamma/\Gamma$ (though the smaller $\lambda \gamma/\Gamma$, the slower the convergence). Such an equality between $I^H_L(t)$ and $\tilde{I}^H_L(t)$ at long times is expected as the energy may be stored only temporarily in the lead-dot coupling region. Finally, we also check that in the long time limit $\Gamma t/\hbar \to \infty$, the t-Kwant particle and heat currents converge to the static particle and heat currents $I^{N/H,\bar{st}}_{L}$ given by the Landauer-B\"uttiker formulas (horizontal dashed lines in Fig.\ref{fig:RLM_TimeDepInDot}), as expected from Eqs.\eqref{eq_INstaticlimit} and \eqref{eq_IHstaticlimit}.

\subsection{Results for a time-dependent voltage bias}
\begin{figure}[h!]
    \centering
    \includegraphics[keepaspectratio,width=0.9\columnwidth]{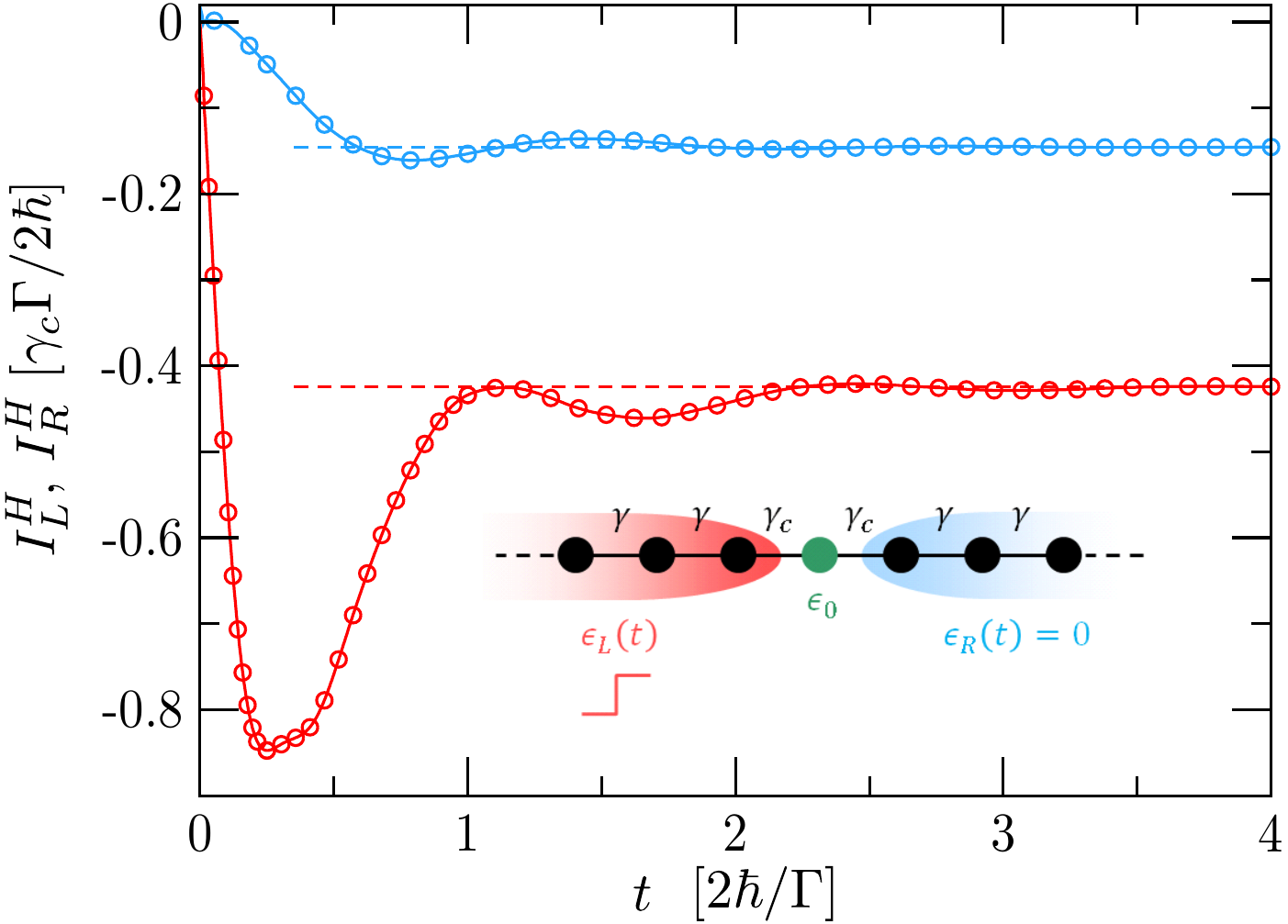}
    \caption{
    Left heat current $I^H_L$ (in red) and right heat current $I^H_R$ (in blue) as a function of time $t$, for the 1D RLM defined by Eqs.\eqref{eq_RLM_1D_1}-\eqref{eq_Hc_RLM}, when a voltage step $\epsilon_L(t)=eV_L\Theta(t)$ is applied in the left lead (sketch in inset). Units are indicated in brackets. The data issued from Ref.\cite{covito2018} (solid lines) and those calculated with t-Kwant (circles) are superimposed. The horizontal dashed lines show the static limits $I^{H,\bar{st}}_{L/R}$ at large times given by the Landauer-B\"uttiker formula (see Sec.\ref{sec:staticlim}). Parameters are fixed to $\epsilon_0=0.2\gamma_c$, $eV_L=2\gamma_c$, $\epsilon_R(t)=0$, $\gamma=5\gamma_c$ $T_L=T_R=0.01\gamma_c/k_B$, and $\mu_L=\mu_R=0$. 
    }
    \label{fig:RLM_TimeDepInLead}
\end{figure}

We continue studying the RLM but now consider that a voltage bias is suddenly applied in the left lead, \textit{i.e.} $\epsilon_L(t)=eV_L\Theta(t)$, while $\epsilon_0(t)=\epsilon_0$ and $\epsilon_R(t)=0$ (see Inset of Fig.\ref{fig:RLM_TimeDepInLead}). This model under the same configuration has been studied in Ref.\cite{covito2018} with an exact (partition-free) numerical approach\cite{eich2016} which is formally equivalent to the t-Kwant approach. The authors calculated the time-dependent particle currents $I^N_\alpha(t)$ in the leads $\alpha=L$ and $R$, as well as some time-dependent heat currents\footnote{$Q_\alpha(t)$ is defined as \unexpanded{$Q_\alpha(t)\equiv -\mathrm{d}\langle \hat{H}_{\alpha}+\frac{1}{2}\hat{H}_{0\alpha}\rangle/\mathrm{d}t$} in Ref.\cite{covito2018} with $\mu_\alpha=0$ throughout the paper.} $Q_\alpha(t)\equiv -\mathrm{d}\langle \hat{H}_{\alpha}+\frac{1}{2}\hat{H}_{0\alpha}\rangle/\mathrm{d}t-\mu_\alpha I^N_\alpha(t)$. In the present case, $Q_\alpha(t)$ and the gauge invariant heat currents $I^H_\alpha(t)$ are linked by the relations (see Appendix \ref{app:heatcurrentdf})
\begin{subequations}
\label{eq_link_Ih_Q_RLM}
\begin{align}
    I^H_L(t)=&\,Q_L(t)-eV_L I^N_L(t)+eV_LN_L\delta(t) \label{eq_link_IhL_QL_RLM}\\
    I^H_R(t)=&\,Q_R(t)
\end{align}
\end{subequations}
where $N_L$ is the particle number in the left lead defined in Eq.\eqref{eq:df_Nalpha}. Note that  $Q_L(t)$ contains the term $-\partial \langle \hat{H}_{L}\rangle/\partial t$ (see Eq.\eqref{eq_ofmotion}) which cancels out the $\delta(t)$ term in Eq.\eqref{eq_link_IhL_QL_RLM}. Using $I^N_\alpha(t)$ and $Q_\alpha(t)$ data\,\footnote{Data are courtesy of Florian Eich. They are the same data as the ones shown in Fig.8 and in the lower panel of Fig.9 of Ref.\cite{covito2018} (for $\lambda=1$).} issued from Ref.\cite{covito2018}, we build up $I^H_\alpha(t)$ data for $t>0$ according to Eq.\eqref{eq_link_Ih_Q_RLM} and compare them to the ones calculated with t-Kwant. We find a perfect agreement (see Fig.\ref{fig:RLM_TimeDepInLead}). This provides a supplemental validity check of our approach and highlights the difference between the gauge invariant heat current $I^H_\alpha(t)$ and the gauge dependent heat current $Q_\alpha(t)$ when a time-dependent voltage is applied in the lead $\alpha$. 

\begin{figure*}[t!]
    \centering
    \includegraphics[keepaspectratio,width=\linewidth]{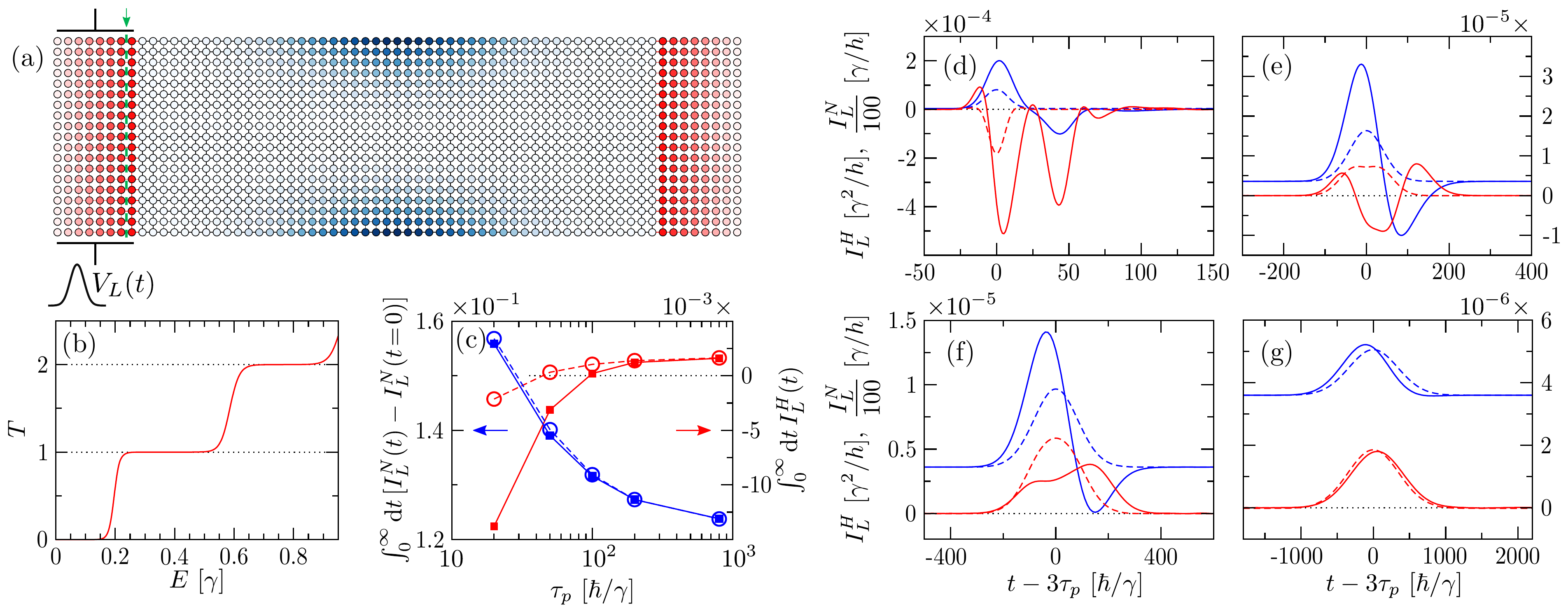}
    \caption{(a) QPC discretized model. The site color in the central region encodes the value of the onsite potential $U_i$ given by Eq.\eqref{eq:QPC_pot} (from 0 (white) to larger values (shades of blue)). A few layers of the left and right semi-infinite leads are shown in red. A voltage pulse $V_L(t)$ is applied in the left lead. Currents are evaluated at the (green dashed) interface indicated by the green arrow. (b) Transmission function $T(E)$ of the QPC defined by $\hat{H}^0$ (see Eq.\eqref{eq:H0_QPC}). (c) $\int_0^\infty\mathrm{d}t\, [I_L^N(t)-I_L^N(t=0)]$ (in blue, in units of $1/2\pi$) and $\int_0^\infty\mathrm{d}t\, I_L^H(t)$ (in red, in units of $\gamma/2\pi$) as a function of the pulse width $\tau_p$ at fixed $n_p=0.2$. Squares with full lines are t-Kwant results, circles with dashed lines are Landauer-B\"uttiker adiabatic results. Lines are guides to the eye. (d) to (g) Left particle currents $I_L^N$ (in blue, in units of $100\gamma/h$) and left heat currents $I_L^H$ (in red, in units of $\gamma^2/h$) as a function of time $t$ (in units of $\hbar/\gamma$), for different widths of the voltage pulse ($\tau_p=20\,\hbar/\gamma$ (d), $100\,\hbar/\gamma$ (e), $200\,\hbar/\gamma$ (f), and $800\,\hbar/\gamma$ (g)) at fixed $n_p=0.2$. Full lines are t-Kwant results, dashed lines are Landauer-B\"uttiker adiabatic results. In all panels, parameters are fixed to $W=18$, $L=48$, $l_x=50$, $l_y=5$, $\mu_L=0.20607\gamma$, $\mu_R=0.2\gamma$, $T_L=0.018\gamma/k_B$, and $T_R=0.02\gamma/k_B$.
    }
    \label{fig:QPC}
\end{figure*}

\section{Quantum Point Contact}
\label{sec:QPC}

To illustrate the potential of our t-Kwant based numerical approach, we simulate hereafter dynamical (electronic) heat transport in a QPC attached to two reservoirs held at different temperatures. We focus on the possibility of extracting heat from the cold reservoir by Peltier effect and ask whether or not Peltier cooling may be enhanced by applying time-resolved voltage pulses to one of the two electrodes attached to the QPC (instead of a constant voltage bias across the system). \\
\indent We consider a nanoribbon of length $L$ and width $W$ connected through semi-infinite leads to two left ($L$) and right ($R$) electronic reservoirs maintained at temperatures $T_L\lesssim T_R$ and electrochemical potentials $\mu_L \gtrsim \mu_R$ (see Fig.\ref{fig:QPC}\,(a)). The system is discretized on a square lattice (with lattice spacing $a=1$). For times $t\leq 0$, no time-dependent perturbation is applied and the system Hamiltonian $\hat{H}(t\leq 0)=\hat{H}^0$ reads
\begin{equation}
    \label{eq:H0_QPC}
    \hat{H}^0=\sum_{i}(4\gamma+U_i)\hat{c}_i^\dagger \hat{c}_i -\gamma \sum_{\langle i,j \rangle}\hat{c}_i^\dagger \hat{c}_j
\end{equation}
where $\gamma$ is the nearest-neighbor hopping term and $U_i$ is the QPC confining potential modeled by
\begin{equation}
    \label{eq:QPC_pot}
    U_i=\left\lbrace
    \begin{array}{l}
    \left(\frac{y_i}{l_y}\right)^2\!\left[1-3\left(\frac{2x_i}{l_x}\right)^2\!\! +2\left|\frac{2x_i}{l_x}\right|^3\right]^2\mathrm{if}~|x_i|<\frac{l_x}{2}\\
   0~~\mathrm{if}~|x_i|\geq\frac{l_x}{2}\,.
    \end{array}
    \right.
\end{equation}
Here $l_x$ and $l_y$ are two parameters controlling the QPC shape and the site of coordinates $(x_i,y_i)=(0,0)$ is taken at the center of the ribbon. The staircase-like transmission function $T(E)$ of the QPC in the static configuration (computed with Kwant) is plotted in Fig.\ref{fig:QPC}(b) for a given set of parameters used hereafter. We also fix $T_L\lesssim T_R$ and choose $\mu_R$ so as $T(E=\mu_R)\approx 0.6$ (guided by the fact that thermoelectric effects are to be sought near transmission steps in the adiabatic regime). The value of $\mu_L \gtrsim \mu_R$ is determined by the condition $I^H_L(t\leq 0)=0$. \\
\indent From time $t=0$, we apply in the left lead a Gaussian voltage pulse $V_L(t)=V_p\exp[-4\ln{2}\frac{(t-3\,\tau_p)^2}{\tau_p^2}]$ of width $\tau_p$, amplitude $V_p$ and center $3\tau_p$. Therefore, the system Hamiltonian becomes $\hat{H}(t> 0)=\hat{H}^0+\sum_{i\in L}V_L(t)\hat{c}_i^\dagger \hat{c}_i$.
Using t-Kwant along with our \texttt{tkwantoperator} extension,\cite{energytKwant} we compute the time-resolved particle ($I^N_L$) and heat ($I^H_L$) currents in the left lead. Data are shown in panels (d) to (f) of Fig.\ref{fig:QPC} for different pulse parameters $(\tau_p,V_p)$ with fixed $n_p\equiv (e/h)\int V_L(t)\mathrm{d}t=(eV_p\tau_p)/(4\hbar\sqrt{\pi\ln2})$ (total number of electrons injected by the voltage pulse in the left lead). To avoid spurious effects that appear when the edges of the system's conduction band are probed,\cite{gaury2014a,gaury2014b} we consider relatively long pulses with $\hbar /\tau_p, V_p \lesssim \mu_L, \mu_R$ (but short enough to investigate the nonadiabatic regime). The t-Kwant currents are compared to the adiabatic currents $I^{N,\bar{st}}_L(V_L(t))$ and $I^{H,\bar{st}}_L(V_L(t))$ given by the Landauer-B\"uttiker formulas (see Sec.\ref{sec:staticlim}). The latter depend parametrically on time through $V_L(t)$. They are computed for static systems by using Kwant and a numerical integrator over the energy. For small $\tau_p$ (short pulses, see panel (d)), the particle current $I^N_L(t)$ shows a first positive peak centered around $3\tau_p$ corresponding to the injected pulse and some time later, a second negative peak corresponding to the reflected part of the pulse. Both peaks are well resolved in this (nonadiabatic) regime. They contribute to two main negative peaks in the heat current $I^H_L(t)$. For large $\tau_p$ (long pulses, see panel (g)), the t-Kwant currents converge to the adiabatic currents characterized by a single peak centered at $3\tau_p$. We note that the particle current converges more slowly to its adiabatic limit than the heat current. The crossover between the two regimes is shown in panels (e) and (f). Obviously, the time-resolved t-Kwant currents in the nonadiabatic regime depend on the position of the interface in the left lead at which they are calculated (green dashed line in Fig.\ref{fig:QPC}\,(a)). However, the currents integrated over time are independent of this position. In panel (c) of Fig.\ref{fig:QPC}, we plot $\int\mathrm{d}t\, [I_L^N(t)-I_L^N(t=0)]$ and $\int\mathrm{d}t\, I_L^H(t)$ as a function of $\tau_p$ ($I_L^H(t=0)=0$ by construction). We find that heat can be extracted from the cold reservoir\footnote{Note that if one substracts the relation $\sum_\alpha [\mu_\alpha I_\alpha^N(t_0)+I_\alpha^H(t_0)]=0$ to Eq.\eqref{eq_1stlaw} and integrates over time, one gets $W_p+W_{ch}+Q_L+Q_R=0$ where $W_p=\int\mathrm{d}t\, [S^E_L(t)+S^E_R(t)+S^E_{\mathcal{S}}(t)]$ (see Eq.\eqref{eq_source_form}), $W_{ch}=(\mu_L-\mu_R)\int\mathrm{d}t\, [I_\alpha^L(t)-I_\alpha^L(t_0)]$, and $Q_\alpha \equiv \int\mathrm{d}t\, [I_\alpha^H(t)-I_\alpha^H(t_0)]$ ($I_L^H(t_0)=0$ in Fig.\ref{fig:QPC}). For all $\tau_p$ shown in Fig.\ref{fig:QPC}(c), $W_p>0$, $W_{ch}>0$, and $Q_R<0$ ($Q_R(t_0)<0$). Peltier cooling of the left reservoir ($Q_L>0$) is only achieved for large $\tau_p$, \textit{i.e.} in the adiabatic regime.} ($\int\mathrm{d}t\, I_L^H(t)>0$) in the limit of long pulses only and for all $\tau_p$, we have $\int\mathrm{d}t\, I_L^H(t)\leq\int\mathrm{d}t\, I^{H,\bar{st}}_L(V_L(t))$. Thus, the application of short voltage pulses involving a nonadiabatic response of the quantum system turns out to be detrimental to Peltier cooling (at least for the set of parameters considered here). A detailed study of Peltier and Seebeck thermoelectric effects in a time-dependent QPC is left for future works. The present preliminary investigation shows the feasibility of further studies. Indeed, the set of t-Kwant curves shown in panels (d) to (g) of Fig.\ref{fig:QPC} required a few hours (d) to a few days (g) of computation time on a single CPU core.

\section{Conclusion}
\label{sec:ccl}
We have built up a gauge invariant theoretical framework for studying time-dependent thermoelectric transport through a broad class of electronic quantum systems, in the absence of electron-electron and electron-phonon interactions. To simulate this approach on a large scale, we have adopted the wave-function formulation of time-dependent quantum transport drawn up in Refs.\cite{gaury2014a,weston2016a,weston2016b}, which is formally equivalent to the NEGF formalism, the Floquet theory (for periodic perturbations), and the partition-free approach. We have thereby implemented a complementary package to the t-Kwant library that allows us to simulate time-dependent energy transport in addition to time-dependent particle transport. We have checked that the built-in platform reproduces the expected results for the time-resolved heat currents in the Resonant Level Model and have performed preliminary investigations of dynamical heat transport in a larger system made up of about one thousand sites.
The approach benefits from t-Kwant advantages in terms of versatility, user-friendliness, and computational efficiency. It provides a numerical test bed for the study of time-dependent thermoelectrics and caloritronics in realistic electronic quantum systems, beyond the adiabatic limit. \\
\indent For the sake of simplicity, we have ignored the spin degree of freedom and considered a spinless model throughout the paper. Yet, there is no technical limitation at Kwant's or t-Kwant's level as both softwares have been devised in such a way that it is easy from a user or a developer point of view to account for spin, orbital, or electron-hole degrees of freedom. The difficulty arises from the choice of the energy operator $\hat{\varepsilon}$ and from the interpretation of the different terms in the energy continuity equation. While for instance the inclusion of the Zeeman term is straightforward, the one of spin-orbit coupling is less obvious.\\
\indent Another natural extension of this work would be the inclusion of the Coulomb interaction at the mean field level. Indeed it was argued by B\"uttiker \textit{et al.}\cite{buttiker1993a,buttiker1993b} that a proper treatment of electrostatics is needed to restore \textit{(i)} particle current conservation \textit{i.e.} $\sum_\alpha I^N_\alpha(t)=0$ and \textit{(ii)} the condition of (strong) gauge invariance \textit{i.e.} the absence of particle current generation upon varying the potential in all the leads simultaneously. This is also true for electronic heat current\cite{chen2015} though \textit{(i)} may not be verified due to dissipation. Importantly, condition \textit{(ii)} is a stronger form of gauge invariance than the one considered in the present paper and it is not satisfied within our noninteracting theory.\footnote{$I^N_\alpha(t)$, $I^E_\alpha(t)$, and $I^H_\alpha(t)$ are invariant under any gauge transformation of the form \eqref{eq_gaugetransfo}, in particular the one which consists in raising the on-site potential everywhere simultaneously. This is not true if such a variation of the potential is made in all the leads but not in the scattering region.}  To go further, one could follow the approach used in Ref.\cite{kloss2018} and solve the time-dependent Hartree problem with t-Kwant in order to describe eventually time-dependent heat and thermoelectric transport together with electrostatic effects.

\acknowledgments
We acknowledge the financial support of the Cross-Disciplinary Program on Numerical Simulation of CEA (the French Alternative Energies and Atomic Energy Commission). We are grateful to Christoph Groth, Thomas Kloss, Beno\^it Rossignol, Xavier Waintal, and Josef Weston for introducing us to t-Kwant and for their help in the implementation of the energy-related routines. We thank Fabienne Michelini for interesting discussions about the energy current definition as well as Florian Eich and his co-authors for sharing the data of Ref.\cite{covito2018} used in Fig.\ref{fig:RLM_TimeDepInLead}.

\appendix

\section{Derivation of the energy continuity equation \eqref{eq_continuityenergy}}
\label{app:proof_eqcontinuity}
Using the equation of motion \eqref{eq_ofmotion} for $\hat{\varepsilon}_i$ and the definition \eqref{eq_kobeoperator} of the energy operator, we find
\begin{equation}
    \frac{\mathrm{d}\rho_i^E}{\mathrm{d}t}-\frac{i}{\hbar}\langle [\hat{\varepsilon}(t),\hat{\varepsilon}_i(t)]\rangle=\frac{i}{\hbar}\langle [e\hat{V}(t),\hat{\varepsilon}_i(t)]\rangle+\langle \frac{\partial \hat{\varepsilon}_i}{\partial t}\rangle\,.
\end{equation}
We identify the right-hand side of the above equation with the source term $S_i^E$ and $(-i/\hbar)\langle [\hat{\varepsilon},\hat{\varepsilon}_i]\rangle$ with $\sum_j I^E_{ji}$. With this choice, the equality $\sum_{i,j} I^E_{ji}=0$ is straightforward. After a few lines of calculation, we get
\begin{align}
    \frac{i}{\hbar}\langle [\hat{\varepsilon},\hat{\varepsilon}_i]\rangle=& \sum_{k,j}\mathrm{Re}\left[ \varepsilon_{ki}\varepsilon_{ij}G^<_{jk}+\varepsilon_{kj}\varepsilon_{ji}G^<_{ik}\right] \label{eq:app_proofEC1}\\
    \frac{i}{\hbar}\langle [e\hat{V},\hat{\varepsilon}_i]\rangle=& \sum_{j}\mathrm{Re}\left[ eV_i\varepsilon_{ij}G^<_{ji}+eV_j\varepsilon_{ji}G^<_{ij}\right] \label{eq:app_proofEC2}\\
    \left\langle \frac{\partial \hat{\varepsilon}_i}{\partial t}\right\rangle=& \sum_j\hbar\,\mathrm{Im}\left[\frac{\partial \varepsilon_{ij}}{\partial t}G_{ji}^<\right] \label{eq:app_proofEC3}\,.
\end{align}
Here the explicit time dependency of the different terms has been omitted for compactness. Eqs.\eqref{eq:app_proofEC2} and \eqref{eq:app_proofEC3} lead to Eq.\eqref{eq_sourceterm} after noticing that the terms for $j=i$ are zero. Eq.\eqref{eq:app_proofEC1} does not allow us to identify immediately the local energy current $I^E_{ji}$ as we seek an expression of $I^E_{ji}$ that satisfies $I^E_{ji}=-I^E_{ij}$. To go further, we use the fact that $\sum_{k,j}\mathrm{Re}[ \varepsilon_{ki}\varepsilon_{ij}G^<_{jk}]=0$ since $\varepsilon_{ij}^*=\varepsilon_{ji}$ and $[G^<_{jk}]^*=-G^<_{kj}$, and we rewrite Eq.\eqref{eq:app_proofEC1} as 
\begin{equation}
    \frac{i}{\hbar}\langle [\hat{\varepsilon},\hat{\varepsilon}_i]\rangle= \sum_{k,j}\mathrm{Re}\left[ \varepsilon_{kj}\varepsilon_{ji}G^<_{ik}-\varepsilon_{ki}\varepsilon_{ij}G^<_{jk} \right]
\end{equation}
to be identified with $-\sum_j I^E_{ji}=\sum_j I^E_{ij}$. This provides Eq.\eqref{IEij_via_G} and completes the proof of Eq.\eqref{eq_continuityenergy}. As pointed out in Sec.\ref{sec:localenergy}, this choice for $I^E_{ji}$ is not unique.

\section{Discussion of the lead heat current definition \eqref{eq_timedepheatcurrent}}
\label{app:heatcurrentdf}

The purpose of this Appendix is to compare the lead heat current $I^H_\alpha(t)$ defined by Eq.\eqref{eq_timedepheatcurrent} with the lead heat current $Q_\alpha(t)$ used in Refs.\cite{ludovico2014,ludovico2016,covito2018} and defined by
\begin{equation}
    \label{eq_app_dfQ}
    Q_\alpha(t)= -\frac{\mathrm{d}}{\mathrm{d}t}\langle \hat{H}_{\alpha}+\frac{1}{2}\hat{H}_{\mathcal{S}\alpha}\rangle -\mu_\alpha I^N_\alpha(t)
\end{equation}
where $\hat{H}_{\alpha}$ is the Hamiltonian of the lead $\mathcal{L}_\alpha$ and $\hat{H}_{\mathcal{S}\alpha}$ the tunneling Hamiltonian between $\mathcal{L}_\alpha$ and the scattering region. We introduce the Hamiltonian density operator
\mathleft
\begin{equation}
    \label{eq_hamiltoniandensity}
    \hat{H}_i(t)=H_{ii}(t)\hat{c}_i^\dagger \hat{c}_i+\frac{1}{2}\sum_{j\neq i}\!\left(H_{ij}(t)\hat{c}_i^\dagger \hat{c}_j+H_{ji}(t)\hat{c}_j^\dagger \hat{c}_i\right)
\end{equation}
\mathcenter
which yields $\sum_{i\in\mathcal{L}_\alpha}\hat{H}_i=\hat{H}_{\alpha}+\frac{1}{2}\hat{H}_{\mathcal{S}\alpha}$. Using $\rho_i^E=\langle \hat{H}_i \rangle-eV_i\rho_i^N$ (which comes from Eq.\eqref{eq_kobeoperator}) and summing over the sites $i$ in $\mathcal{L}_\alpha$, we find with the help of Eqs.\eqref{eq:df_Ealpha} and \eqref{eq_timedepheatcurrent}
\begin{equation}
\label{eq_app_IHvsQ}
    I^H_\alpha(t)=Q_\alpha(t)+eN_\alpha(t)\frac{\partial V_\alpha}{\partial t}-eV_\alpha(t) I^N_\alpha(t)\,.
\end{equation}
$I^H_\alpha(t)$ is gauge invariant while the three terms on the right-hand side of Eq.\eqref{eq_app_IHvsQ} are not. However in the t-Kwant gauge \eqref{eq:df_tkwantgauge} in which the leads are time-independent, $I^H_\alpha(t)=\widetilde{Q}_\alpha(t)$ where $\widetilde{Q}_\alpha(t)$ is defined by Eq.\eqref{eq_app_dfQ} upon replacing $\hat{H}$ by the gauge-transformed Hamiltonian $\hat{\widetilde{H}}$ given in Sec.\ref{sec_tkwantgauge}. Thereby we recover Eq.\eqref{eq_timedepheatcurrent_vs_lit}.\\
\indent As a side note, let us add that the counterpart of the continuity equation \eqref{eq_continuityenergy} for the Hamiltonian density reads 
\begin{equation}
    \label{eq_app_continuityeqforH}
    \frac{\mathrm{d} \langle  \hat{H}_i \rangle}{\mathrm{d}t}+\sum_{j\neq i}I^{H}_{ji}(t)=\left\langle \frac{\partial  \hat{H}_i}{\partial t} \right\rangle
\end{equation}
where 
\begin{align}
    I^{H}_{ji}(t)=\sum_k \mathrm{Re}[H_{kj}H_{ji}G_{ik}^< 
    -H_{ki}H_{ij}G_{jk}^<]
\end{align} 
(dropping the explicit time dependence of the different terms). None of the three terms in Eq.\eqref{eq_app_continuityeqforH} is in general gauge invariant, contrary to the ones of Eq.\eqref{eq_continuityenergy}.

\section{Convergence to the static limit}
\label{app:static}
We assume that the Hamiltonian $\hat{H}(t)$ defined in Sec.\ref{sec:model} converges to a static limit $\hat{H}(t\to\infty)=\hat{H}^{\bar{st}}$ at long times. We derive Eq.\eqref{eq_statcurrent2} with the help of Eqs.\eqref{eq:IN(t)_via_WF} and \eqref{eq:IE(t)_via_WF} upon omitting the tildes in those equations since the proof given below does not require working in the t-Kwant gauge \eqref{eq:df_tkwantgauge}.

\subsection{Particle current}
We prove Eq.\eqref{eq_INstaticlimit} in two steps. We focus first on the static problem defined by $\hat{H}^{\bar{st}}$ for all times. The local particle current for this static problem is given by Eq.\eqref{eq:IN(t)_via_WF}
\begin{align}
    \label{eq_app_IN_ji_st}
    I_{ji}^{N,\bar{st}}=2& \sum_\beta\sum_{m_\beta} \int\frac{\mathrm{d}E}{h} f_{\mu_\beta+eV_\beta,T_\beta}(E) \nonumber \\
    & \times \mathrm{Im}\left[ \left(\Psi_j^{m_\beta E,\bar{st}}\right)^* H_{ji}^{\bar{st}} \,\Psi_i^{m_\beta E,\bar{st}}\right]
\end{align}
where $\Psi_i^{m_\beta E,\bar{st}}$ is the stationary scattering state at site $i$ corresponding to an incoming mode $m_\beta$ in lead $\mathcal{L}_\beta$ with energy $E$ \textit{i.e}
\begin{equation}
    \label{eq_app_statSchrodinger}
    \mathbf{H}^{\bar{st}}\Psi^{m_\beta E,\bar{st}}=E\Psi^{m_\beta E,\bar{st}}\,.
\end{equation}
Note that the static electric potential $V_\beta$ is included in the leads (\textit{i.e} $H_{ii}^{\bar{st}}=H_{ii}^0+eV_\beta$ if $i\in\mathcal{L}_\beta$) and in the reservoirs through the Fermi-Dirac distribution. We now make use of the periodic pattern of each semi-infinite lead built of identical unit cells, labeled $x=1,2,...$ from the scattering region. In the stationary case, the total particle current $I_\alpha^{N,\bar{st}}$ in the lead $\mathcal{L}_\alpha$ (given by Eqs.\eqref{eq_IN_lead} and \eqref{eq_app_IN_ji_st}) is invariant along the lead axis and we have for any $x$
\begin{align}
\label{eq_app_INst_1}
    I_\alpha^{N,\bar{st}}=-2& \sum_\beta\sum_{m_\beta} \int\frac{\mathrm{d}E}{h} f_{\mu_\beta+eV_\beta,T_\beta}(E) \nonumber \\
    & \times \mathrm{Im}\left[ \left(\Psi_{\alpha,x-1}^{m_\beta E,\bar{st}}\right)^\dagger W_\alpha \,\Psi_{\alpha,x}^{m_\beta E,\bar{st}}\right]
\end{align}
$\Psi_{\alpha,x}^{m_\beta E,\bar{st}}$ being the scattering state in the $x$-th cell of the lead $\mathcal{L}_\alpha$ corresponding to an incoming mode $m_\beta$ in lead $\mathcal{L}_\beta$ with energy $E$ and $W_\alpha$ the coupling matrix connecting neighboring unit cells in $\mathcal{L}_\alpha$. Using the notations of Ref.\cite{gaury2014a}, we write the scattering state $\Psi_{\alpha,x}^{m_\beta E,\bar{st}}$ as a superposition of plane waves  
\begin{align}
    \label{eq_app_PsiInLead}
    \Psi_{\alpha,x}^{m_\beta E,\bar{st}}=&\,\delta_{\alpha\beta}\frac{\xi_{m_\beta}^{in}}{ \sqrt{\hbar|v^{in}_{m_\beta}|}}e^{-ik^{in}_{m_\beta}x} \nonumber \\
    & + \sum_{m_\alpha}\frac{\xi_{m_\alpha}^{out}}{\sqrt{\hbar|v^{out}_{ m_\alpha}|}}e^{ik^{out}_{m_\alpha}x}S_{m_\alpha m_\beta}^{\alpha\beta}
    \end{align}
where the sum runs over the modes $m_\alpha$ in lead $\mathcal{L}_\alpha$. The vectors $\xi_{\alpha, m_\alpha}^{in}(E)$ and $\xi_{\alpha,m_\alpha}^{out}(E)$ defined on one unit cell are the transverse parts of the incoming and outgoing modes $m_\alpha$ with energy $E$ in lead $\mathcal{L}_\alpha$. $k^{in}_{m_\alpha}(E)$, $k^{out}_{m_\alpha}(E)$, and $v^{in}_{\alpha, m_\alpha}(E)$, $v^{out}_{\alpha, m_\alpha}(E)$ are the corresponding mode momenta and velocities. $S_{m_\alpha m_\beta}^{\alpha\beta}(E)$ is the scattering amplitude of an electron injected at energy $E$ from the lead $\mathcal{L}_\beta$ in mode $m_\beta$ into the mode $m_\alpha$ in lead $\mathcal{L}_\alpha$. By inserting Eq.\eqref{eq_app_PsiInLead} into Eq.\eqref{eq_app_INst_1} and by using the relations\cite{wimmer2009,gaury2014a}
\begin{align}
    &i(\xi_{m_\alpha}^{in})^\dagger[e^{-ik^{in}_{n_\alpha}}W_\alpha-e^{ik^{in}_{m_\alpha}}W_\alpha^\dagger]\xi_{n_\alpha}^{in}=\delta_{n_\alpha m_\alpha}\hbar v^{in}_{m_\alpha} \label{eq_app_orthomodes_1}\\
    &i(\xi_{m_\alpha}^{out})^\dagger[e^{ik^{out}_{n_\alpha}}W_\alpha-e^{-ik^{out}_{m_\alpha}}W_\alpha^\dagger]\xi_{n_\alpha}^{out}=\delta_{n_\alpha m_\alpha}\hbar v^{out}_{m_\alpha} \label{eq_app_orthomodes_2}\\
    &i(\xi_{m_\alpha}^{out})^\dagger[e^{-ik^{in}_{n_\alpha}}W_\alpha-e^{-ik^{out}_{m_\alpha}}W_\alpha^\dagger]\xi_{n_\alpha}^{in}=0 \label{eq_app_orthomodes_3}
\end{align}
it can be shown that $I_\alpha^{N,\bar{st}}$ reduces to the standard Landauer-B\"uttiker formula
\begin{equation}
   \label{eq_app_IN_st_LB}
   I_\alpha^{N,\bar{st}} =
  \sum_{\beta \neq \alpha}\!\int\!\frac{\mathrm{d}E}{h}\,\bar{T}_{\alpha\beta}(E) [f_{\mu_\alpha+eV_\alpha,T_\alpha}(E)-f_{\mu_\beta+eV_\beta,T_\beta}(E)]   
\end{equation}
where $\bar{T}_{\alpha\beta}=\sum_{m_\alpha}\sum_{m_\beta}|S_{m_\alpha,m_\beta}^{\alpha\beta}|^2$.

Let us now consider the time-dependent problem defined by $\hat{H}(t)$. In that case, the local particle current $I_{ji}^{N}(t)$ given by Eq.\eqref{eq:IN(t)_via_WF} reads
\begin{align}
 I^N_{ji}(t) = 2 & \sum\limits_\beta \sum_{m_\beta} \int\frac{\dint E}{h} f_{\mu_\beta,T_\beta}(E) \nonumber \\
 & \times\imag\left[\left(\Psi^{m_\beta E}_j(t)\right)^* H_{ji}(t) \Psi^{m_\beta E}_i(t)\right] .
   \label{eq_app_IN_ji_t}
\end{align}
To calculate $I^N_{ji}(t\to\infty)$ using $\hat{H}(t\to\infty)=\hat{H}^{\bar{st}}$, it is important to notice first that $E$ in the equation above labels the energy of an incoming mode $m_\beta$ in lead $\mathcal{L}_\beta$ in the remote past \textit{i.e} for $t\leq t_0$. In that case, the on-site potential in $\mathcal{L}_\beta$ is $H_{ii}(t\leq t_0)=H_{ii}^0$ while $H_{ii}^{\bar{st}}=H_{ii}^0+eV_\beta$ (if $i\in\mathcal{L}_\beta)$.
For this reason, 
\begin{equation}
    \label{eq_app_wflimit}
    e^{i\theta_E(t)}\,\Psi_j^{m_\beta E}(t) \xrightarrow[t\to\infty]{} \Psi_j^{m_\beta, E+eV_\beta,\bar{st}}
\end{equation}
where $\theta_E(t)$ is an (irrelevant) spatially constant phase. \commentout{Actually, even if it seems reasonable, I am not able to prove it mathematically. I did some numerical tests in the t-Kwant gauge (of course) and found
\begin{equation}
    e^{iEt/\hbar}e^{ie(\phi_\beta(t)-\Lambda_j(t))/\hbar}\,\widetilde{\Psi}_j^{m_\beta E}(t) \xrightarrow[t\to\infty]{} \Psi_j^{m_\beta, E+eV_\beta,\bar{st}}
\end{equation}
(where $\Lambda_j(t)$ is defined by Eq.\eqref{eq:df_tkwantgauge}) while when I try to (semi-)prove it, I rather find
\begin{equation}
    e^{i(E+eV_\beta)t/\hbar}e^{-ie\Lambda_j(t)/\hbar}\,\widetilde{\Psi}_j^{m_\beta E}(t) \xrightarrow[t\to\infty]{} \Psi_j^{m_\beta, E+eV_\beta,\bar{st}}\,.
\end{equation}
In both cases, I have the form \eqref{eq_app_wflimit} (switching to the natural gauge) but it is not $100\%$ understood.} Doing the change of variable $E'=E+eV_\beta$ in Eq.\eqref{eq_app_IN_ji_t} and comparing with Eq.\eqref{eq_app_IN_ji_st}, we find $I_{ji}^{N}(t\to\infty)=I_{ji}^{N,\bar{st}}$. We deduce Eq.\eqref{eq_INstaticlimit} by using Eqs.\eqref{eq_IN_lead} and \eqref{eq_app_IN_st_LB}.

\subsection{Energy current}

Let us consider first the static problem defined by $\hat{H}^{\bar{st}}$ for all time. For this static problem, we define the energy operator as $\hat{\varepsilon}=\hat{H}^{\bar{st}}$. With this definition, the local energy current given by Eq.\eqref{eq:IE(t)_via_WF} simplifies to
\begin{align}
    \label{eq_app_IE_ji_st}
    I_{ji}^{E, \bar{st}}=& \, 2\sum_\beta\sum_{m_\beta} \int\frac{\mathrm{d}E}{h} f_{\mu_\beta+eV_\beta,T_\beta}(E)E\nonumber \\
    & \times \mathrm{Im}\left[ \left(\Psi_j^{m_\beta E, \bar{st}}\right)^* H_{ji}^{\bar{st}} \,\Psi_i^{m_\beta E, \bar{st}}\right]
\end{align}
after using the static Schr\"odinger equation \eqref{eq_app_statSchrodinger}. To calculate the energy current $I^{E,\bar{st}}_\alpha$ in the lead $\mathcal{L}_\alpha$ with Eq.\eqref{eq_timedepenergycurrent} , it is convenient to redefine the scattering region $\mathcal{S}$ -- as we did previously to write down Eq.\eqref{eq_app_INst_1} -- by including into it the first cell $x=1$ of $\mathcal{L}_\alpha$ (or an arbitrary number of cells $x=1,2,...)$. This does not change $I^{E,\bar{st}}_\alpha$ as the static energy current calculated with $\hat{\varepsilon}=\hat{H}^{\bar{st}}$ is invariant along the lead axis. By using Eqs.\eqref{eq_app_IE_ji_st} and \eqref{eq_app_PsiInLead}-\eqref{eq_app_orthomodes_3}, we find
\begin{align}
   I_\alpha^{E,\bar{st}} =
  \sum_{\beta \neq \alpha}&\int \frac{\mathrm{d}E}{h}\,E\,\bar{T}_{\alpha\beta}(E) \nonumber \\
  & \times [f_{\mu_\alpha+eV_\alpha,T_\alpha}(E)-f_{\mu_\beta+eV_\beta,T_\beta}(E)]\,. 
  \label{eq_app_IE_st_LB}
\end{align}
Note that Eq.\eqref{eq_app_IE_st_LB} is the usual Landauer-B\"uttiker formula for the lead energy current in the static case, which we recovered upon defining in this case the energy operator as $\hat{\varepsilon}=\hat{H}^{\bar{st}}$.

Let us consider on the other hand the time-dependent problem defined by $\hat{H}(t)$. The energy operator is now defined by Eq.\eqref{eq_kobeoperator}. Using Eqs.\eqref{eq:IE(t)_via_WF}, \eqref{eq_app_wflimit}, and finally \eqref{eq_app_statSchrodinger}, we find for the local energy currents in the long time limit
\begin{equation}
    \label{eq_app_IE_ji_t_bis}
    I_{ji}^{E}(t\to\infty)=I_{ji}^{E, \bar{st}}-\frac{e}{2}(V_i+V_j)I_{ji}^{N,\bar{st}} 
\end{equation}
where $V_i\equiv V_i(t\to\infty)$ is a shorthand notation for the long time limit of the external time-dependent scalar potentials $V_i(t)$. We deduce from Eq.\eqref{eq_timedepenergycurrent} 
\begin{equation}
  I^E_\alpha(t\to\infty)=I_{\alpha}^{E,\bar{st}} -eV_\alpha  I_{\alpha}^{N,\bar{st}} +S^E_\alpha(t\to\infty)
\end{equation}
since $V_i=V_\alpha$ if $i\,\in\,\mathcal{L}_\alpha$  and $\frac{\partial \phi_{ij}}{\partial t}\to 0$ in the static limit $t\to\infty$. This concludes the proof of Eq.\eqref{eq_IEstaticlimit}. 

\commentout{
\subsection{Contribution of the lead-scatterer tunneling Hamiltonian to the energy current}
We show hereafter that $\frac{\mathrm{d}\langle \hat{H}_{\mathcal{S}\alpha}\rangle}{\mathrm{d}t}\to 0$ in the static limit $t\to \infty$, $\hat{H}_{\mathcal{S}\alpha}$ being the tunneling Hamiltonian between the lead $\mathcal{L}_\alpha$ and the scattering region $\mathcal{S}$. The proof goes as follows. Using the equation of motion \eqref{eq_ofmotion}, we get $\frac{\mathrm{d}\langle \hat{H}_{\mathcal{S}\alpha}\rangle}{\mathrm{d}t}\to (i/\hbar)[\hat{H}^{\bar{st}}, \hat{H}^{\bar{st}}_{\mathcal{S}\alpha}]$ when $t\to \infty$. 
\begin{align}
    \frac{i}{\hbar}[\hat{H}^{\bar{st}}, \hat{H}^{\bar{st}}_{\mathcal{S}\alpha}]=& \sum_i \sum_{p\in\mathcal{S}}\sum_{q\in\mathcal{L}_\alpha}\left\lbrace H_{ip}^{\bar{st}}H_{pq}^{\bar{st}}G^<_{qi}-H_{pq}^{\bar{st}}H_{qi}^{\bar{st}}G^<_{ip}\right. \nonumber\\
    &\left.+H_{iq}^{\bar{st}}H_{qp}^{\bar{st}}G^<_{pi}-H_{qp}^{\bar{st}}H_{pi}^{\bar{st}}G^<_{iq}\right\rbrace
\end{align}
}

\section{Resonant Level Model within the\\
Non Equilibrium Green's Function formalism}
\label{app_RLM_NEGF}
In this appendix, we give the RLM formula for the lead particle current $I^{N}_\alpha(t)$ and the lead heat currents $I^{H}_\alpha(t)$, $\tilde{I}^{H}_\alpha(t)$ that are used in Fig.\ref{fig:RLM_TimeDepInDot} to plot the NEGF curves. The model under consideration is the one introduced in Sec.\ref{sec:RLM_Model} with $\epsilon_0(t)=\epsilon_0+eV_0\Theta(t)$ and $\epsilon_L(t)=\epsilon_R(t)=0$. The lead Hamiltonians $\hat{H}_\alpha$ and the tunneling Hamiltonians between the dot and the leads $\hat{H}_{0\alpha}$ are written in the reciprocal space, as 
\begin{align}
    \hat{H}_\alpha= &\sum_{k_\alpha}\epsilon_{k_\alpha}\hat{c}^\dagger_{k_\alpha}\hat{c}_{k_\alpha}\\
    \hat{H}_{0\alpha}= &\sum_{k_\alpha}V_{k_\alpha}\hat{c}^\dagger_{k_\alpha}\hat{c}_{0}+h.c.
\end{align}
where $\hat{c}_{k_\alpha}=\sum_{j\in\alpha}e^{i j k_\alpha }\hat{c}_j$ is the annihilation operator of an electron with momentum $k_\alpha$ in lead $\alpha=L$ or $R$, $V_{k_\alpha}=\gamma_c \,\mathrm{sin}(k_\alpha)$ the hybridization term, and $\epsilon_{k_\alpha}=-2\gamma\, \mathrm{cos}(k_\alpha)$ the dispersion relation (with a lattice spacing fixed to unity). Then the currents are calculated within the NEGF formalism under the wide-band limit approximation, \textit{i.e} assuming that $\Gamma_\alpha(E)\equiv -2\,\imag \Sigma^R(E)=2\pi\sum_{k_\alpha}|V_{k_\alpha}|^2\delta(E-\epsilon_{k_\alpha})$ is energy independent ($\Gamma_L=\Gamma_R=\Gamma/2$). This is true in the limit $\lambda\gamma/\Gamma\gg 1$ as noticed in Sec.\ref{sec:RLM_timeindot}. We refer to the seminal paper\cite{jauho1994} of Jauho \textit{et al.} for the derivation of the particle current and to Refs.\cite{crepieux2011,zhou2015,dare2016,ludovico2016} for its extension to the energy and heat currents. We gather here the results.
Introducing the notations $\hat{K}_\alpha^N=\hat{N}_\alpha=\sum_{i\in\alpha}\hat{c}^\dagger_i\hat{c}_i$, $\hat{K}_\alpha^E=\hat{H}_\alpha+\frac{1}{2}\hat{H}_{0\alpha}$, and $\hat{K}_\alpha^{\tilde{E}}=\hat{H}_\alpha$, we have for $\lambda=N$, $E$ and $\tilde{E}$
\begin{equation}
    \label{eq_app_RLM_integral}
    \left\langle \frac{\mathrm{d}\hat{K}_\alpha^\lambda}{\mathrm{d}t}\right\rangle =\sum_\beta \int\frac{\mathrm{d}E}{2\pi}f_\beta(E)\, \mathcal{I}^{\lambda}_{\alpha\beta}(E,t)
\end{equation}
where the sum over $\beta$ is made over both leads $L$ and $R$, and
\mathleft
\begin{align}
    \mathcal{I}^N_{\alpha\beta}(E,t)=\, &\frac{\Gamma}{\hbar}\left[\frac{\Gamma}{4}|A(E,t)|^2+\delta_{\alpha\beta}\mathrm{Im}A(E,t) \right]\\
    \mathcal{I}^{\tilde{E}}_{\alpha\beta}(E,t)=\, &E\,\mathcal{I}^N_{\alpha\beta}(E,t)\! +  \!\frac{\Gamma^2}{4}\mathrm{Im}[A(E,t)\frac{\partial A^*}{\partial t}(E,\!t)] \\
    \mathcal{I}^E_{\alpha\beta}(E,t)=\, & \mathcal{I}^{\tilde{E}}_{\alpha\beta}(E,t) + \frac{\Gamma}{2}\delta_{\alpha\beta}\mathrm{Re}\left[\frac{\partial A}{\partial t}(E,t)\right]
\end{align}
\mathcenter
while the spectral density $A(E,t)$ reads
\begin{equation}
    A(E,t)=\frac{E-\epsilon_0+i\frac{\Gamma}{2}-eV_0e^{i(E-\epsilon_0-eV_0+i\frac{\Gamma}{2})t/\hbar}}{(E-\epsilon_0+i\frac{\Gamma}{2})(E-\epsilon_0-eV_0+i\frac{\Gamma}{2})}\,.
\end{equation}
We used the formula above to plot the NEGF particle current $I^N_\alpha(t)=-\langle \frac{\mathrm{d}\hat{N}_\alpha}{\mathrm{d}t} \rangle$ and the NEGF heat currents $I^H_\alpha(t)=-[ \langle \frac{\mathrm{d}\hat{K}_\alpha^E}{\mathrm{d}t} \rangle - \mu_\alpha \langle \frac{\mathrm{d}\hat{N}_\alpha}{\mathrm{d}t} \rangle]$ and $\tilde{I}^H_\alpha(t)=-[ \langle \frac{\mathrm{d}\hat{K}_\alpha^{\tilde{E}}}{\mathrm{d}t} \rangle - \mu_\alpha \langle \frac{\mathrm{d}\hat{N}_\alpha}{\mathrm{d}t}\rangle]$ in Fig.\eqref{fig:RLM_TimeDepInDot} (circles).
The integrals over the energy were computed numerically.

\section{t-Kwant extension package for energy transport: overview and performance}
\label{app:performance}
To calculate our newly defined energy related quantities, we have implemented a Python package, \texttt{tkwantoperator},\cite{energytKwant} as an extension to t-Kwant,\cite{tKwant} with additional classes : \texttt{EnergyDensity}, \texttt{EnergySource} and \texttt{EnergyCurrent\-Divergence} can be called for calculating respectively $\rho_i^E$, $S_i^E$, and $\sum_{j\neq i}I^E_{ji}$ over a given list of sites $\{i\}$; \texttt{EnergyCurrent} for calculating the current $I^E_{ji}$ flowing through a given list of hoppings between sites $\{(j,i)\}$;  \texttt{Lead\-Heat\-Current} for calculating the heat current $I^H_\alpha$ in a given lead $\mathcal{L}_\alpha$. Since the classes that are called to evaluate the particle and energy operators all have a similar structure, another class \footnote{we implemented additionally a general operator Cython class that can be used to calculate quantities of the form 
\begin{equation*}
\sum\limits_{(i_0, \dots, i_{n+1})} \phi_{i_0}^\dagger O^{(0)}_{i_0 i_1} O^{(1)}_{i_1 i_2} \dots O^{(n)}_{i_n i_{n+1}} \psi_{i_{n+1}} \pm \text{c.c.}
\label{eq:genOp}
\end{equation*}
where $\phi(t)$ and $\psi(t)$ are wave functions calculated by t-Kwant. $O^{(0)}(t), \dots, O^{(n)}(t)$ and $(i_0, \dots, i_{n+1})$ are respectively quadratic operators and tuple of sites in $\mathcal{S}$ that are to be defined when using the class.\commentout{The particle and energy operators defined in this paper are particular cases of this general class.} Subclasses inheriting from this general class can be easily implemented to define \textit{e.g.} the particle and energy operators or other operators enabling the computation of correlation functions. It comes however at the cost of an unfriendly implementation of the general class and of (slightly) longer computation times (when used for the particle and energy operators, in comparison to the direct implementation).} generalizing the former ones has also been implemented but not yet used. \commentout{More information about the practical use of these classes can be found in the online documentation.}

The calculation of the many-body expectations values of the various operators involves an integration over the energy $E$ and a sum over the modes $m_\alpha$ injected at this energy from all the leads $\mathcal{L_\alpha}$. Since the resolution of the Schr\"odinger-like differential equation \eqref{eq:tdepSE3} giving the evolution in time of the scattering states $\Psi^{m_\alpha E}(t)$ is the most time-consuming task of the t-Kwant algorithm (see below), it is crucial to use as few scattering states as possible to evaluate the expectations values. For this purpose, a Gauss-Kronrod adaptive scheme\cite{weston2016b} is used when integrating the contribution of each state over the energy. It determines the needed number $N_\mathrm{scat}$ of scattering states for a given precision on the expectation values. Moreover, the time evolution of the scattering states can be done in parallel on multi-core computers, each core dealing with a subset of the scattering states. Both functionalities implemented in the core version of t-Kwant are leveraged to compute the expectation values of the energy operators.

\begin{figure}
    \centering
    \includegraphics[keepaspectratio,width=\columnwidth]{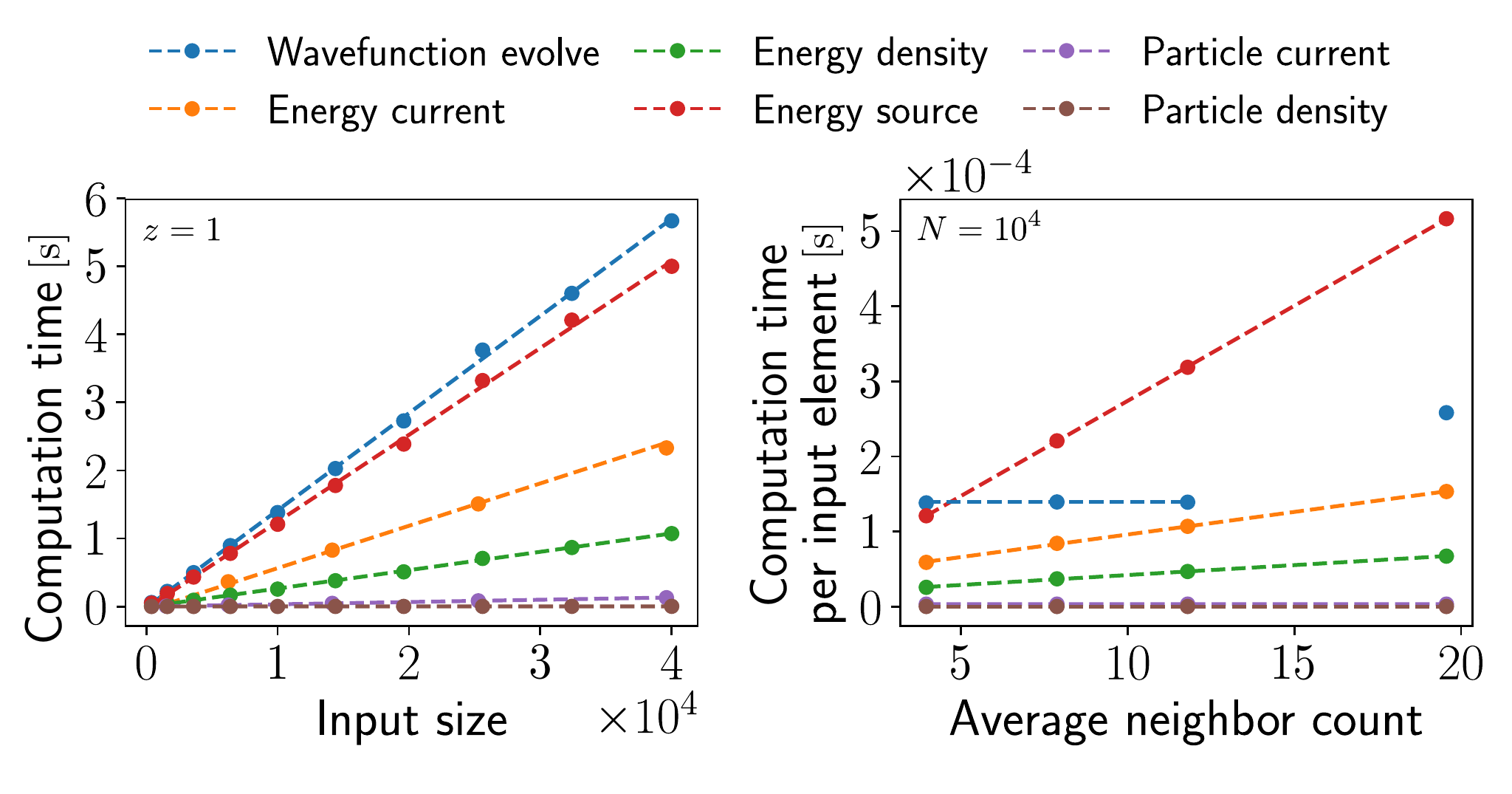}
    \caption{Comparison of the computation times needed for the evolution of a single wave function by a time step and for the evaluation of its contribution to the particle and energy operators. Data (bullets) are shown for the square system made of $N=L^2$ sites defined in the text. Its Hamiltonian includes hopping terms up to the $z$-th nearest neighbors. Dashed lines are linear fits. (Left) CPU times for evaluating operators and evolving the wave function, as a function of the size of their input site/hopping tuples (varied by increasing $L$, for fixed $z=1$). The input size equals $N=L^2$ for the wave function and the density/source operators, while it equals the number of hoppings $M_{[z=1]}/2=L(L-1)$ for the current operators. (Right) CPU times divided by the input size $N$ or $M_z/2$, as a function of the average number of neighbors per site $M_z/N$, varied by taking $z=1,2,3,$ and $4$ at fixed $N=10^4$ sites. \commentout{The used scripts are available online.}}
    \label{fig:speed-test}
\end{figure}

Hereafter, we analyze the extra CPU time cost due to the evaluation of the energy operators. Given that the computation times for evolving the scattering states (between times $t_{n-1}$ and $t_n$) and for calculating a many-body expectation value (at a time $t_n$) grow linearly with the total number $N_\mathrm{scat}$ of scattering states, we compare these two computation times for only one wave function $\Psi$. The wave function is initialized (at $t_0=0$) with uniformly distributed random complex values on each system site, in the $[-1, 1] \times [-\iu, \iu]$ complex square. The computation time for the stationary problem, done once for a given system, is not considered here. Investigations of the t-Kwant CPU times are done for a closed (\textit{i.e.} without leads) square system with $N=L^2$ sites lying on a square lattice. The onsite potential $H_{ii}$ is disordered and shifted by a time-dependent perturbation for $t\geq t_1$ as
\begin{align}
    H_{ii}(t> 0) = w_i + \Theta(t-t_1)&[\,\mathrm{sin} (\alpha t) \, \mathrm{e}^{-\beta t^2} \nonumber\\
    & + \eta(1 + \mathrm{tanh} (\delta t))]
\end{align}
where $t_1=0.8, \alpha=8, \beta=15, \eta=0.3, \delta=10$, and $w_i$ are random values that are normally distributed around zero with a standard deviation of $0.025$. Hopping terms between sites $i\neq j$ are fixed to $H_{ij}=\gamma(=1)$ up to the $z$-th nearest neighbors and are zero beyond. Each site $i$ thus has $M_i^z$ connected neighbors. We note $M_z=\sum_i M_i^z$. In Fig.\ref{fig:speed-test}, we compare the computation times used for making the wave function $\Psi(t)$ evolve by a time step and for calculating its contribution to the various particle and energy operators. Its contribution reads for instance $\sum_j\mathrm{Re}[\Psi_i^*\epsilon_{ij}\Psi_j]$ for the energy density operator evaluated on site $i$ (see Eq.\eqref{eq:rho_E_via_WF}). Each point in Fig.\ref{fig:speed-test} is obtained by averaging the computation times of 200 measurements performed at times $t_n$ (or over the intervals $[t_n,t_{n+1}]$ for the evolution of $\Psi(t)$) evenly spaced between $t_0=0$ and $t_{max}=2$. CPU times are expressed in seconds and result from simulations run on a single core (Intel Xeon Silver 4114 CPU at 2.20GHz, 32 GB RAM).\\
\indent We check on the left panel of Fig.\ref{fig:speed-test} \textit{(i)} that the CPU time used for evolving a wave function by a time step grows linearly with the number of sites $N$ (as already reported in Refs.\cite{gaury2014a, weston2016a}), and \textit{(ii)} that the CPU times corresponding to the computation of the contributions to the various operators grow linearly with the size of the lists of sites or hoppings on which they are calculated. The relative positions of the straight lines in this panel (obtained for $z=1$) show us that it takes (much) longer to calculate the energy operators than the particle ones (which is obvious in view of the mathematical expression of the operators) but that the global CPU time used by the simulation is nevertheless dominated by the calculation of the wave-function evolution. In the right panel of Fig.\ref{fig:speed-test}, we investigate how this picture is modified when second ($z=2$), third ($z=3$), and fourth ($z=4$) nearest-neighbors are included. The CPU time used for the wave-function evolution is unaffected (except for $z=4$ due to unknown -- probably memory -- reasons), as well as the CPU time corresponding to the particle density and the CPU time per hopping corresponding to the particle current. On the contrary, the CPU times corresponding to the energy operators are much increased since their expressions involve a sum over neighboring sites. 
It is to be noted at that stage that often, in practice, the operators only need to be calculated on a subsystem while the wave function has to be calculated necessarily over the $N$ sites of the system. For instance, the lead (particle, energy, heat) currents are calculated at the interface between the leads and the scattering region which involves a negligible number of hoppings in comparison to the total number $M_z/2$ of hoppings in the system. For this reason, we conclude that evaluating operators has a low-to-negligible impact on the global t-Kwant computation time for most practical situations. For completeness, let us add that the CPU times needed for evaluating the operators and evolving the scattering states depend at a quantitative level on the simulated systems and on the hardware used. Additional (not shown) data indicate that this should not affect qualitatively the conclusion given above.

\bibliography{biblio}

\end{document}